\documentclass[sn-mathphys-num]{sn-jnl}

\usepackage{amsmath,amssymb,amsfonts}%
\usepackage{amsthm}%
\usepackage{xcolor}%
\usepackage{algorithm}%
\usepackage{algpseudocode}%

\usepackage{mathtools}

\renewcommand{\figurename}{Figure}

\usepackage{xr}

\usepackage{subfiles} 

\usepackage{fge}

\usepackage{bbm}%

\newcommand{\given}{\,|\,}

\newcommand{\figwa}{0.625}
\newcommand{\figwb}{0.9}
\newcommand{\figlong}{0.8}
\newcommand{\figwide}{0.99}

\newcommand{\mysetminus}{\mathbin{\fgebackslash}}

\newif\ifshowgraphics
\showgraphicstrue
\newcommand{\maybeincludegraphics}[2][]{%
    \ifshowgraphics
        \includegraphics[#1]{#2}%
    \fi
}

\newif\ifusesvg
\usesvgfalse
\newcommand{\figext}{%
    \ifusesvg
        svg%
    \else
        png%
    \fi
}

\begin{document}

\title[]{Bayesian adaptive design for neural recruitment curves}

\author*[1]{\fnm{Vishweshwar} \sur{Tyagi}}\email{vt2353@cumc.columbia.edu}
\author[1]{\fnm{Saihari S.} \sur{Dukkipati}}
\author[2]{\fnm{Daniel} \sur{Twili}}
\author[2]{\fnm{Jonas M.} \sur{Mikhaeil}}
\author[2]{\fnm{Andrew} \sur{Gelman}}
\author[1]{\fnm{Jason B.} \sur{Carmel}}
\equalcont{Joint senior authors.}
\author*[1]{\fnm{James R.} \sur{McIntosh}}\email{jrm2263@cumc.columbia.edu}
\equalcont{Joint senior authors.}
\affil*[1]{\orgdiv{Department of Neurology}, \orgname{Columbia University}, \orgaddress{\city{New York}, \state{NY}, \postcode{10032}}}
\affil[2]{\orgdiv{Department of Statistics}, \orgname{Columbia University}, \orgaddress{\city{New York}, \state{NY}, \postcode{10027}}}

\abstract{
    Nonlinear response models remain ubiquitous across disciplines with recruitment curves describing muscle response to neural stimulation.
    Conventional acquisition methods typically rely on static uniform sampling, requiring many stimuli, while adaptive threshold-hunting designs such as ML-PEST forfeit accurate characterization of the overall curve.
    We present a Bayesian adaptive design that uses posterior uncertainty over curve parameters to propose stimuli that maximize information gain across the full recruitment curve.
    We additionally implement a focused acquisition method that can target arbitrary subset of curve parameters, which we use to evaluate and interpret our design.
    In simulations, our design outperforms uniform sampling in recovering threshold while matching ML-PEST, and outperforms ML-PEST in predictive curve accuracy while matching uniform.
    It likewise matches our threshold-focused acquisition, suggesting our design doesn't trade off threshold estimation for recovering the curve.
    We further show data collected using our design yields greater power when detecting a change in threshold in intervention studies.
    Our design proposes the next stimulus within 5 seconds, making it feasible for transcranial magnetic stimulation (TMS) experiments.
    Finally, we validate our design in TMS experiments and provide a library for Python, called \texttt{onMEP}, for online acquisition of neural recruitment curves.
}


\maketitle

\section{Introduction}\label{intro}
\begin{figure}[tph!]
    \centering
    \maybeincludegraphics[width=\figlong\textwidth]{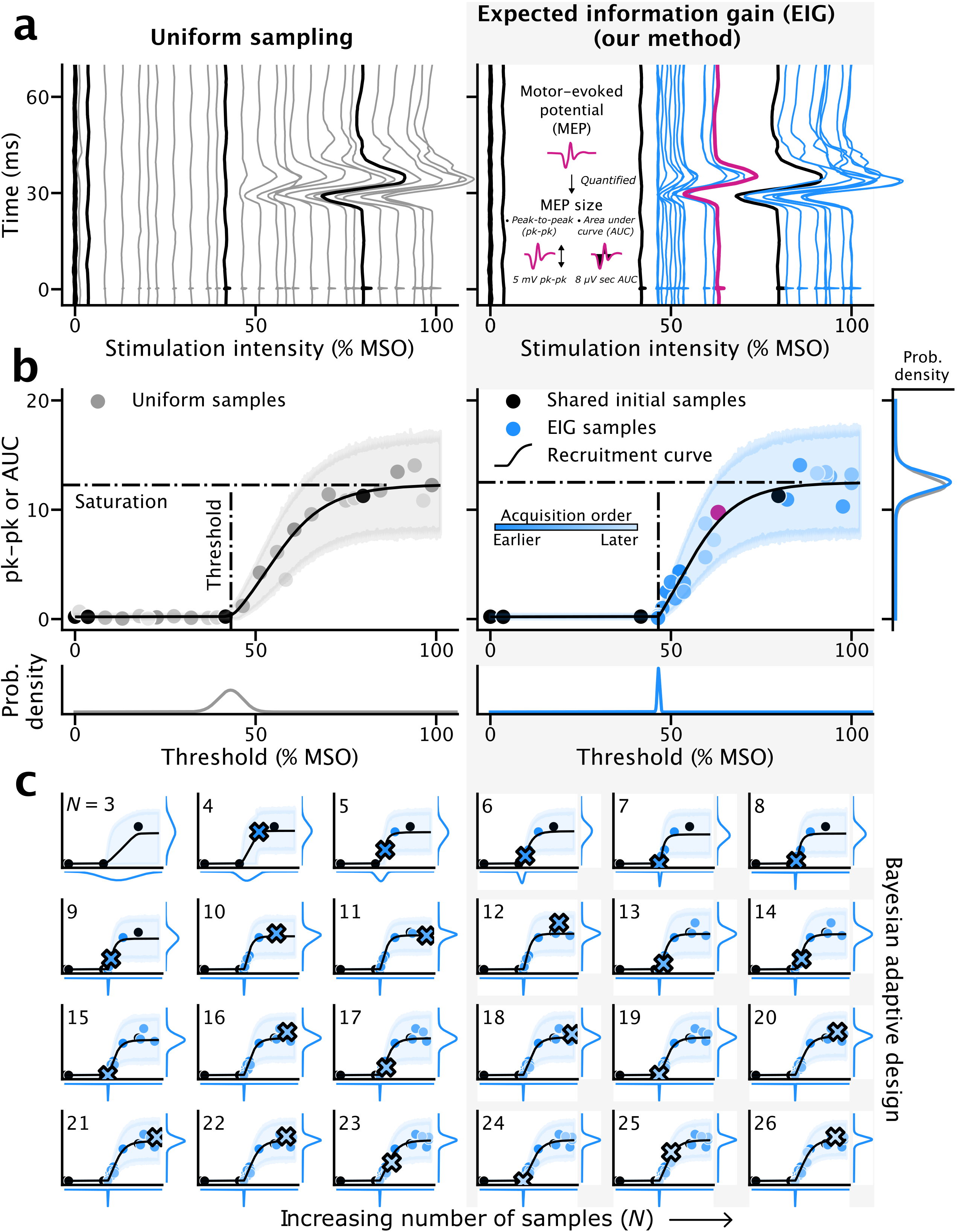}
    \caption{
    \textbf{Bayesian adaptive acquisition of recruitment curves using posterior uncertainty over curve parameters to maximize information gain.}
    \textbf{(a)} Example MEPs from a real participant's APB muscle in TMS experiment, recorded at uniform intensities (left), and at intensities proposed by the expected information gain (EIG) design (right).
    Both uniform and EIG are block randomized and start with the same initial samples (black dots)---eight samples at zero intensity (not counted) and three uniform samples.
    The abscissa is specified as percent maximum stimulator output (\% MSO) for TMS but may instead use units of current, such as $\mu$A or mA, for spinal cord stimulation.
    Inset: schematic quantification of MEPs using either peak-to-peak (pk-pk) amplitude or area under the curve (AUC).
    \textbf{(b)} Recruitment curves estimated from uniform and EIG samples at the end of the experiment, with 26 samples each. 
    Samples are colored by acquisition order from earlier (dark) to later (light).
    EIG yields a narrower threshold posterior (bottom), indicating greater certainty in threshold estimate, and allocates samples to more informative regions of the curve.
    Side panel: posterior distribution of saturation.
    \textbf{(c)} Starting with three initial samples, the EIG design updates the posterior over curve parameters and uses it to propose the intensity for the next sample (cross mark).
    It samples near the threshold early and later samples the transition regions to estimate the slope and saturation.
    Accordingly, the threshold posterior (bottom) narrows early, while the saturation posterior (right) narrows later.
}
\label{fig_intro}
\end{figure}
Neural recruitment curves characterize nonlinear muscle response to neural stimulation.
They are derived using electromagnetic and electrical stimulation techniques, such as transcranial magnetic stimulation (TMS), spinal cord stimulation (SCS), and nerve stimulation, which are used to assess and modulate the neural systems that control movement \citep{stefan_induction_2000,chen_clinical_2008,picht_preoperative_2011,millet_electrical_2011,bunday_motor_2012,macdonald_intraoperative_2013,sayenko_spinal_2015,hofstoetter_spinal_2021,greiner_recruitment_2021,pal_spinal_2022,arora_utility_2022,mcintosh_intraoperative_2023,balbinot_segmental_2023,koponen_transcranial_2024}.
When targeted to the sensorimotor system, these techniques produce motor-evoked potentials (MEPs) (Fig.~\ref{fig_intro}a), which are quantified by their size, typically in terms of peak-to-peak amplitude or area under the curve (Fig.~\ref{fig_intro}b, gray and blue dots).
The recruitment curve (Fig.~\ref{fig_intro}b, black curve) models the MEP size as a non-linear, non-decreasing function of stimulation intensity and exhibits a characteristic S-shape, with a steep increase above the threshold leading up to a plateau phase at high intensities (Fig.~\ref{fig_intro}b, lower and side panels).
Whereas parameters such as the slope and saturation may have specific neurophysiological interpretations, the threshold remains a critical parameter for inferring changes in corticospinal excitability \citep{devanne_input-output_1997,farzan_single-pulse_2014,schicktanz_motor_2014,vucic_clinical_2023,bestmann_use_2024}, while the overall recruitment curve characterizes the total corticospinal output over the stimulation range \citep{carson_characterizing_2013,potter-baker_influence_2016,kraus_recruitment_2018,iyer_characterization_2019,kemlin_redundancy_2019,filipovic_neurophysiological_2021,eginyan_effects_2023,murphy_lower_2023}.

In this paper, we consider the design of experiments for acquiring recruitment curves.
Specifically, we consider designs that sequentially determine the stimulation intensity at which to acquire the next MEP sample.
Collecting data using TMS and SCS can be a costly and burdensome process for both the participant and the experimenter \citep{antal_no_2004,kukke_efficient_2014,alavi_optimal_2019,hassanzahraee_longer_2019,manson_relationship_2020,mcintosh_intraoperative_2023,bestmann_use_2024,sharma_repeated_2024}, and our requirements for an efficient design are therefore high accuracy in recovering both the threshold and the overall curve using a limited number of stimuli, typically fewer than 40.

Conventional methods for recruitment curve acquisition include static designs that use uniform sampling, which requires many stimuli for accurate threshold estimation and results in samples located in non-informative regions of the curve, such as well below the threshold (Fig.~\ref{fig_intro}b, left).
Ideally, we would reallocate these samples to more informative regions along the rising portion of the curve, such as near the threshold and in the upper region transitioning to saturation.
Moreover, such designs lack a continuous feedback loop during acquisition and require the experimenter to predetermine the number of stimuli to be tested.
Consequently, experiments can neither terminate early when the curve has been estimated with sufficient certainty nor continue longer when additional samples are needed in the presence of high measurement noise.
Other designs exist, most notably the adaptive threshold-hunting design known as maximum likelihood parameter estimation by sequential testing (ML-PEST) \citep{taylor_pest_1967,pentland_maximum_1980,harvey_efficient_1997,awiszus_tms_2003,mishory_maximum-likelihood_2004,wang_three_2023}.
However, ML-PEST is designed specifically for estimating the resting motor threshold and consequently forfeits accurate characterization of the overall curve.

In contrast, Bayesian experimental design provides a principled information-theoretic framework for dynamically adapting experimental decisions as data is acquired \citep{lindley_measure_1956,chaloner1995bayesian,rainforth_nesting_2018,rainforth_modern_2023}.
We introduce a Bayesian adaptive design that uses posterior uncertainty over curve parameters to propose stimuli that maximize the expected information gain (EIG) across the recruitment curve (Fig.~\ref{fig_intro}b, right).
It concentrates samples in informative regions of the curve, focusing on the threshold in the early phase and subsequently sampling the transition regions to estimate the slope and saturation (Fig.~\ref{fig_intro}c).
We use a generative hierarchical framework to simulate high-fidelity synthetic data that closely matches empirical TMS observations for model comparison \citep{tyagi_hierarchical_2025}.
In simulations, we compare the number of stimuli needed to reach acceptable error in recovering the motor threshold and the recruitment curve.
We additionally implement a focused version of our method that can target information gain in an arbitrary subset of curve parameters, which we use to evaluate and interpret our design.
To ensure practical feasibility, we optimize our design to reduce the time to choose the next stimulus and evaluate its speed and accuracy when acquiring either a single target muscle or multiple muscles simultaneously.
For a use case, we show data collected using our design yields greater statistical power and requires fewer participants to detect a change in the threshold in intervention studies.
Finally, we validate our design in human TMS experiments and provide a library for Python, called \texttt{onMEP}, for online acquisition of neural recruitment curves.

\section{Results}\label{results}
\begin{figure}[tph!]
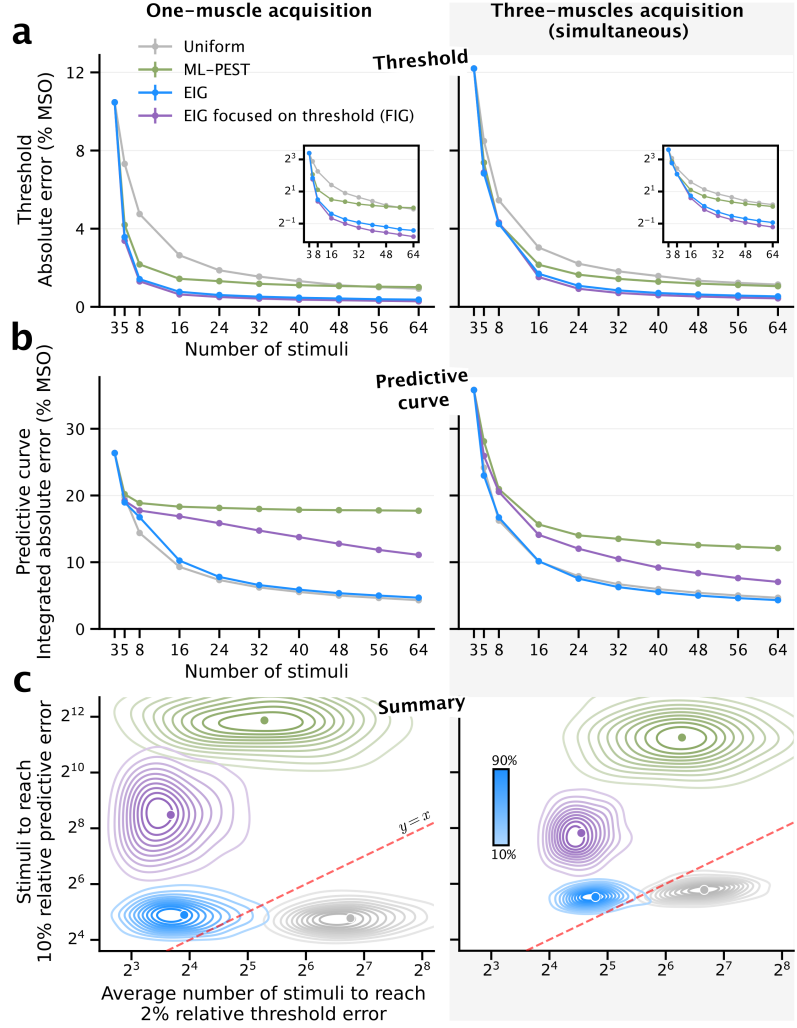

    \centering
    \maybeincludegraphics[width=\figlong\textwidth]{pdf/acc.\figext}
    \caption{
    \textbf{EIG simultaneously improves threshold and recruitment curve estimation on sparse data.}
    \textbf{(a)} Mean absolute error in threshold estimation as a function of number of stimuli when acquiring a single target muscle (left) and three muscles simultaneously, with error averaged across muscles (right).
    Error bars represent standard error of the mean.
    Inset: zoom to show error on logarithmic scale.
    \textbf{(b)} Mean integrated absolute error for the estimated predictive curve.
    \textbf{(c)} Joint distribution of the average number of stimuli needed to reach relative threshold error of $2$\% and relative predictive error of $10$\%, estimated from experiments with $16$ participants.
    Contours span the (10, 90)\% iso-proportion interval, and midpoints indicate the mean.
    The dashed diagonal line indicates equal numbers of stimuli needed to reach both error criteria simultaneously.
    }
\label{fig_accuracy}
\end{figure}
\subsection{Accurate estimation on sparse data}\label{res_sparse}
We previously introduced a Bayesian model for neural recruitment curves \citep{tyagi_hierarchical_2025}, which we extended into a Bayesian adaptive design that sequentially selects stimulation intensities based on the data acquired during experiment (Methods~\ref{meth_bad}).
The design selects the next intensity by maximizing the expected information gain (EIG), which measures how much a new observation is expected to reduce uncertainty about the recruitment curve.
We additionally implemented focused EIG that can target information gain in an arbitrary subset of curve parameters, allowing the design to target a specific parameter, such as the threshold, or the full curve, depending on the choice of subset.
To enable a comparative analysis of acquisition methods, we used a hierarchical model estimated on 14 uninjured TMS participants to simulate synthetic data that closely matched empirical observations \citep{tyagi_hierarchical_2025}.
We evaluated EIG for its accuracy in recovering the curve parameters and compared it against three methods---the conventional methods of uniform sampling (U) and ML-PEST (ML), as well as against EIG focused on the threshold (Focused Information Gain or FIG).
In particular, we used each design to acquire data for a given recruitment curve and subsequently used our Bayesian model to estimate the curve parameters from the acquired data.
By keeping the estimation model fixed, we could discriminate between the designs based on their acquired data (Methods~\ref{meth_eval}).

EIG demonstrated improved accuracy in recovering both the threshold and the curve when acquiring a single target muscle (Fig.~\ref{fig_accuracy}a,b, left panels).
Beginning with the same three initial samples, the threshold error was identical across designs.
By $16$ stimuli, EIG achieved lower threshold error than uniform ($e_\text{EIG} - e_\text{U} = -1.87 \pm 0.03$) and ML ($-0.67 \pm 0.06$).
FIG had the lowest error, although it was comparable to EIG ($e_\text{EIG} - e_\text{FIG} = 0.13 \pm 0.02 \in (-0.5, 0.5)$\% MSO).
The improvement over uniform was largest at this relatively low number of stimuli, with the gap narrowing at $64$ stimuli, albeit still not comparable ($e_\text{EIG} - e_\text{U} = -0.56 \pm 0.02 \notin (-0.5, 0.5)$\% MSO).
EIG likewise improved recovery of the overall curve and achieved predictive error comparable to uniform at $32$ stimuli ($0.36 \pm 0.05 \in (-0.5, 0.5)$), whereas ML ($-11.41 \pm 0.17$) performed the worst, followed by FIG ($-8.17 \pm 0.14$),
illustrating the trade-off between acquiring for a single parameter and learning the full recruitment curve.

For most practical use cases, EIG minimized the number of stimuli needed to jointly reach acceptable accuracy in threshold and curve estimation (Fig.~\ref{fig_accuracy}c, left panel).
We estimated the number of stimuli required on average to reach a relative threshold error of $2$\% and relative predictive error of $10$\% by sampling experiments with $16$ participants, a sample size representative of typical studies (Methods~\ref{meth_eval}).
EIG matched ML and FIG in the number of stimuli required for threshold recovery, as indicated by their overlap along the threshold axis, while matching uniform in curve recovery, indicated by their overlap along the predictive axis.
Additionally, during early acquisition, stimuli proposed by EIG matched those of FIG, with their acquisition functions having similar shapes and peaking around the same stimulus values close to threshold (Supplementary Fig.~\ref{sufig_traj}).
Thus, EIG is comparable to designs focused on threshold while retaining accurate recovery of the overall curve, and choosing a different design would require more stimuli to reach either criterion.
Note in the limiting case of a large number of participants, we expect these distributions to shrink towards their mean values (the midpoints), consistent with the comparison shown above (Fig.~\ref{fig_accuracy}a,b).

\subsection{Extension to multiple muscles}\label{res_multiple}
In TMS and SCS experiments, MEPs are often recorded from multiple muscles simultaneously, as opposed to a single target muscle \citep{suzuki_reciprocal_2012,greiner_recruitment_2021,nazarova_mapping_2021,yuasa_systematic_2022,mcintosh_intraoperative_2023,koponen_transcranial_2024,murray_timing-dependent_2025,pascual-leone_electrode_2026,de_freitas_spinal_2026}.
In these cases, an efficient design must select the next stimulus by optimizing acquisition over multiple muscles.
EIG naturally combines information across different muscles into a single acquisition function to jointly maximize information gain (see Methods~\ref{meth_extension}), even when the muscles differ in response scales, as can occur with differences in cortical representation of muscles in TMS or electrode placement in SCS.
Since ML-PEST has no analogous acquisition function, we instead paired it with a round-robin strategy across muscles.

The relative advantages of EIG were preserved when acquiring three muscles simultaneously, with EIG achieving lower average error across muscles for both threshold and curve estimation (Fig.~\ref{fig_accuracy}a--c, right panels).
As expected, the number of stimuli required to reach acceptable threshold error increased across all designs because acquisition now spanned multiple muscles.
In contrast, simultaneous acquisition improved curve estimation for the focused designs. This is because when muscles have sufficiently separate thresholds, samples allocated near the threshold of one muscle can fall within the recruitment region of another, thereby providing additional information about its curve as opposed to single muscle acquisition.
This also benefits EIG during early acquisition, when samples are concentrated around the threshold, as evident from EIG and uniform achieving comparable predictive error at a relatively low number of $16$ stimuli ($e_\text{EIG} - e_\text{U} = 0.05 \pm 0.06 \in (-0.5, 0.5)$).
Nonetheless, EIG continued to minimize the number of stimuli required to jointly reach acceptable error in threshold and curve estimation (Fig.~\ref{fig_accuracy}c, right panel).

\subsection{Online acquisition feasibility}\label{res_online}
\begin{figure}[tph!]
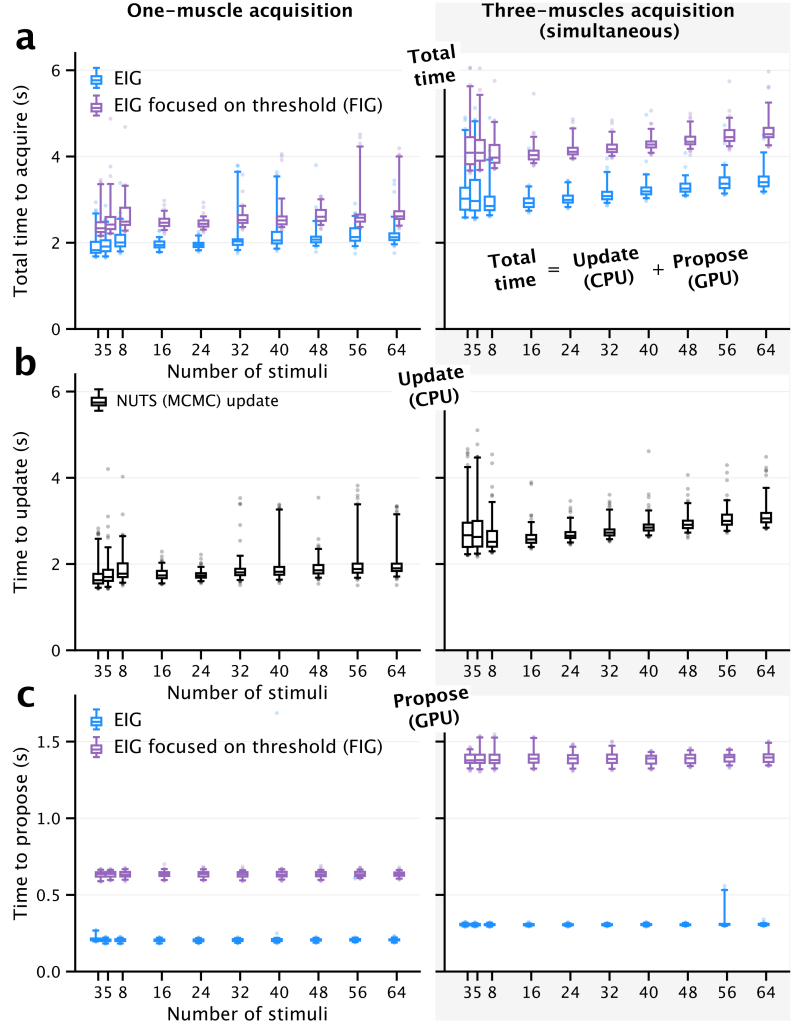

    \centering
    \maybeincludegraphics[width=\figlong\textwidth]{pdf/time.\figext}
    \caption{
    \textbf{EIG proposes the next stimulus within 5 seconds, making it feasible for online TMS experiments.}
    \textbf{(a)} Total computation time between consecutive stimuli for EIG and EIG focused on threshold when acquiring a single target muscle (left) and three muscles simultaneously (right).
    Boxes indicate the median time and span the interquartile range, whiskers span the (2.5, 97.5) percentiles, and dots indicate outliers.
    \textbf{(b)} Computation time for the posterior update step using the No-U-Turn sampler (NUTS), which accounts for most of the total computation time and gradually increases with the number of stimuli.
    \textbf{(c)} Computation time for the proposal step for EIG and EIG focused on threshold, which remains constant with the number of stimuli.
    }
\label{fig_timings}
\end{figure}
EIG consists of two computational steps between consecutive stimuli---the update step, which estimates the posterior distribution over the curve parameters after collecting a new observation, and the proposal step, which computes the acquisition function over candidate stimuli to choose the next stimulus.
The update step uses Markov chain Monte Carlo (MCMC) methods, such as the No-U-Turn sampler (NUTS) \citep{hoffman_no-u-turn_2011}, whereas the acquisition function can be computed across candidate stimuli in parallel.
To account for this, we implemented posterior updating on the CPU and acquisition evaluation on the GPU within our Python library \texttt{onMEP}.
For multiple-muscle acquisition, the update and proposal steps are further parallelized across muscles, such that, with sufficient compute, the required time should approach that of a single muscle.

EIG proposes the next stimulus within $5$ seconds, making it feasible for online TMS experiments (Fig.~\ref{fig_timings}).
For one-muscle acquisition, the total computation time for EIG remained below $3$ seconds, with posterior updating accounting for most of the computation time (Fig.~\ref{fig_timings}a--c, left panels).
The update time increases gradually as observations accumulate, whereas proposal time remains constant because the posterior dimension and candidate stimuli remain fixed.
For simultaneous three-muscle acquisition, parallelization across muscles limits the total time to below $4$ seconds despite tripling the number of recruitment curves being acquired (Fig.~\ref{fig_timings}a--c, right panels).
Thus, both EIG and focused EIG remain below $5$ seconds throughout acquisition in either case, with focused EIG requiring more time due to its costlier acquisition function.
These computation times fall well within the typical interstimulus interval of at least $5$ seconds used in TMS experiments \citep{julkunen_does_2012,hassanzahraee_longer_2019,weise_precise_2023,hernandez-pavon_tms_2023}.

\subsection{Detecting a shift in threshold}\label{res_shift}
\begin{figure}[t]
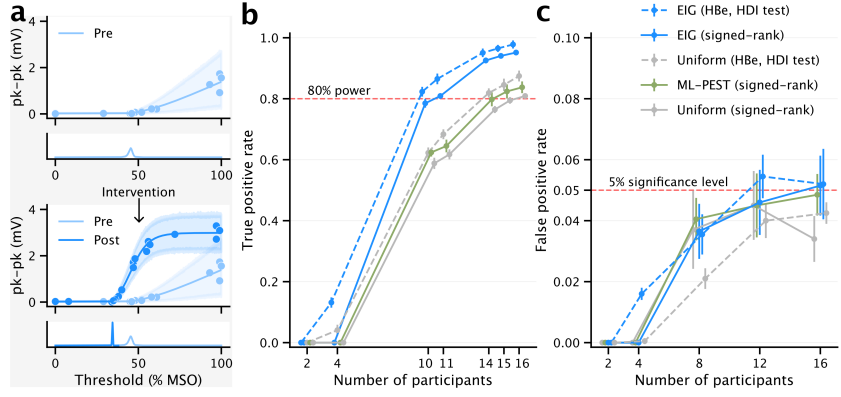

    \centering
    \maybeincludegraphics[width=.85\textwidth]{pdf/power.\figext}
    \caption{
    \textbf{EIG increases statistical power and requires fewer participants when detecting a shift in threshold.}
    \textbf{(a)} Example pre- and post-intervention recruitment curves with the intervention resulting in a negative shift in threshold.
    Bottom panel: threshold posterior.
    \textbf{(b)} True positive rate as a function of the number of participants.
    Here the threshold differences are simulated from a $\mathrm{normal}\!\left(\mu\!=\!-3, \sigma\!=\!2.5\right)$ distribution, where the alternative hypothesis (non-zero shift) is true.
    \textbf{(c)} Comparison of false positive rates against the set significance level of 5\%.
    The differences are simulated from $\mathrm{normal}\!\left(0, 2.5\right)$, where the null hypothesis (zero shift) is true.
    All simulations consisted of 16 samples each for the pre- and post-intervention phases, not counting eight samples at zero intensity.
    Error bars represent the standard error of the mean rate.
    }
\label{fig_power}
\end{figure}
Changes in threshold are used to assess changes in corticospinal excitability in intervention studies \citep{devanne_input-output_1997,farzan_single-pulse_2014,schicktanz_motor_2014,vucic_clinical_2023,bestmann_use_2024}.
We evaluated how the acquisition designs perform in this case by simulating two hypothetical scenarios---an intervention resulting in a negative shift in threshold, indicating increased excitability (Fig.~\ref{fig_power}a), and an intervention resulting in no shift.
We used each design to acquire data for the pre- and post-intervention recruitment curves and subsequently tested for a shift in threshold using a paired two-sided signed-rank test on the point threshold estimates at the $5$\% significance level.
We additionally used hierarchical Bayesian estimation (HBe), which directly models participant-level differences in threshold between the pre- and post-intervention phases and summarizes these differences across participants using a population-level location parameter \citep{tyagi_hierarchical_2025}, with its 95\% highest density interval (HDI) used for hypothesis testing \citep{gelman_type_2000,kruschke_bayesian_2013,kruschke_doing_2014,vincent_hierarchical_2016,tyagi_hierarchical_2025}.
For HBe, the decision rule rejects the null hypothesis (zero shift) in favor of the alternative (non-zero shift) if the 95\% HDI excludes zero.
This allowed us to evaluate whether the non-uniform samples acquired by EIG would affect HBe inference either by reducing statistical power or inflating the false positive rate.

EIG paired with Bayesian estimation required the fewest participants to detect a shift in threshold  (Fig.~\ref{fig_power}b,c).
Using the signed-rank test, uniform sampling required $16$ participants to reach $80$\% power (80.9 $\pm$ 0.9), compared with $15$ for ML-PEST (82.4 $\pm$ 2.3; 79.8 $\pm$ 2.3 with 14 participants), and only $11$ for EIG (80.9 $\pm$ 0.8), a $26.7$\% reduction over ML-PEST.
HBe further increased statistical power with uniform sampling requiring $14$ participants (81.9 $\pm$ 1.5), one less than ML-PEST paired with the signed-rank test.
Finally, EIG paired with HBe required only $10$ participants (82.3 $\pm$ 1.6), albeit with a smaller improvement over the signed-rank test, likely because EIG samples already enable accurate threshold estimation at the participant level.
Furthermore, the false positive rates did not exceed the $5$\% significance level for any method, validating that the differences in statistical power were not because of an inflated false positive rate.

\subsection{Validation in TMS experiments}\label{res_val}
\begin{figure}[tph!]
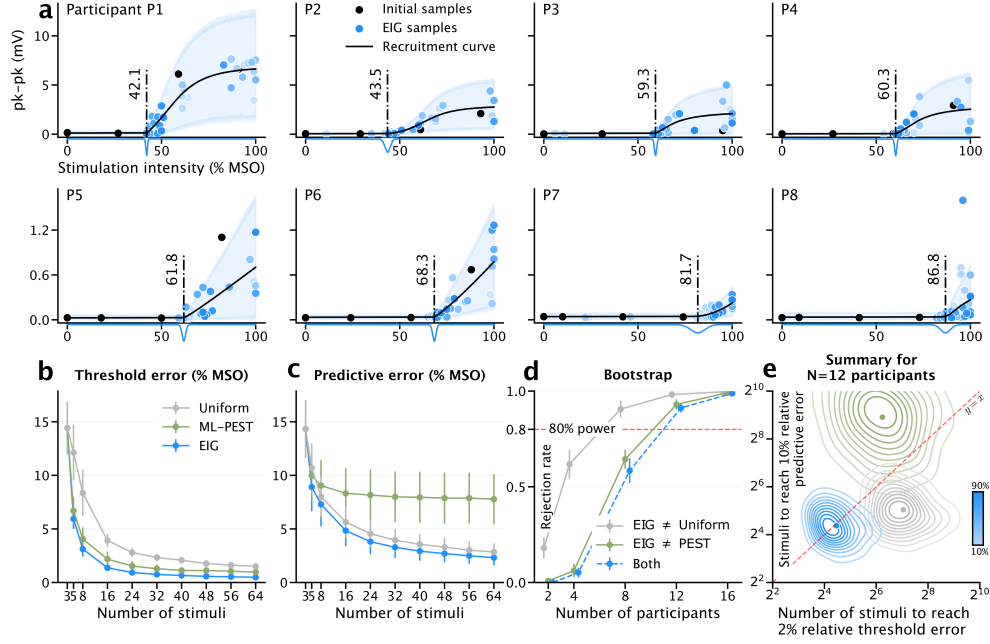

    \centering
    \maybeincludegraphics[width=\figwide\textwidth]{pdf/val.\figext}
    \caption{
    \textbf{Validation of EIG in real TMS experiments.}
    \textbf{(a)} Recruitment curves acquired using EIG from APB muscle of eight participants enrolled in TMS study, with 40 stimuli each, not counting eight samples at zero intensity.
    Samples are colored by acquisition order from earlier (dark) to later (light).
    Bottom: threshold posterior.
    Participants are ordered by increasing threshold.
    \textbf{(b)} Mean absolute error in threshold estimation.
    Error bars represent the standard error across participants.
    \textbf{(c)} Mean integrated absolute error in recruitment curve estimation.
    \textbf{(d)} Rejection rates for the null hypotheses of no difference between EIG and uniform sampling in the number of stimuli needed to reach 2\% relative threshold error (gray curve), between EIG and ML-PEST in stimuli needed to reach 10\% relative predictive error (green curve), and for both hypotheses simultaneously (dashed blue curve).
    \textbf{(e)} Joint distribution of the number of stimuli needed to reach acceptable threshold and predictive errors with $N\!=\!12$ participants.
    Contours span the (10, 90)\% iso-proportion interval, and midpoints indicate the mean.
    The dashed diagonal line indicates equal numbers of stimuli needed to reach both error criteria simultaneously.
    }
\label{fig_val}
\end{figure}
Figure~\ref{fig_val}a shows data acquired using EIG from eight neurotypical participants without evidence of neurological injury enrolled in an ongoing TMS study.
The early adaptive samples (dark blue) are concentrated around the threshold, with later samples (light blue) extending across the recruitment curve.
For participants P7 and P8, with high thresholds exceeding $80$\% MSO, EIG could still locate their thresholds despite all initial samples (black dots) falling within the baseline response, and subsequently allocate samples along the rising portion of their recruitment curves, even with the limited stimulation range available beyond $80$\% MSO.
This is desirable when tuning stimulation parameters, as it can quickly identify configurations for which the threshold lies beyond the available stimulation range and may therefore require adjustment.
In contrast, uniform sampling would result in most samples within the baseline response, providing little information about either their thresholds or the rising portion of the curves.

To determine a sample size for a study comparing the acquisition methods, we performed a prospective power analysis by bootstrapping data from the eight participants (Fig.~\ref{fig_val}b--e).
Because data were collected only using EIG, we used our model to estimate a generative distribution for each participant and sampled from it to evaluate each acquisition method (Methods~\ref{meth_eval}).
Consistent with our earlier results, EIG improved both threshold and recruitment curve estimation (Fig.~\ref{fig_val}b,c).
Based on our simulations, we defined two primary null hypotheses---EIG not differing from uniform sampling in the number of stimuli needed for threshold estimation (Fig.~\ref{fig_val}d, gray curve), and not differing from ML-PEST in stimuli needed for curve estimation (green curve).
The study was considered successful if both null hypotheses were rejected simultaneously (dashed blue curve), which reached $80\%$ power with 12 participants.
At this sample size, the joint distributions of the number of stimuli needed to reach acceptable errors were consistent with our simulations, with EIG achieving the minimum (Fig.~\ref{fig_val}e).

\section{Discussion}\label{discussion}
We introduce a Bayesian adaptive design that uses posterior uncertainty over recruitment curve parameters to maximize expected information gain (EIG).
It improves threshold estimation over uniform sampling while matching ML-PEST, and improves predictive accuracy over ML-PEST while matching uniform.
These advantages extend to simultaneous acquisition over multiple muscles.
We show EIG performs comparably to EIG focused on threshold; hence, it can be adopted without compromising on threshold estimation while retaining predictive accuracy.
For a use case, we show data collected using EIG yields greater statistical power when detecting a change in threshold in intervention studies.
We further optimize our design to operate within the 5-second interstimulus interval for TMS and demonstrate its feasibility in acquiring real data.
Finally, our Python library \texttt{onMEP} for online acquisition of neural recruitment curves makes our approach broadly accessible to researchers for diverse applications.

EIG improves recruitment curve acquisition without trading off threshold estimation.
Our primary motivation in developing the design was to improve upon uniform sampling, rather than methods that target a single curve parameter.
In general, recruitment curves provide a more robust characterization of corticospinal output than a single parameter or MEPs recorded at a single intensity \citep{devanne_input-output_1997,kukke_efficient_2014}.
They use data from multiple trials across a stimulation range and yield, in addition to threshold, other curve parameters including slope and saturation, which are subsequently analyzed offline.
As such, uniform sampling is a natural choice for characterizing the overall curve.
However, it is inefficient because it allocates samples well below threshold, where they provide little information about the curve.
EIG instead reallocates these samples to more informative regions along the rising portion of the curve (Fig.~\ref{fig_intro},~\ref{fig_val}a).
However, this improvement in acquisition efficiency should not come at the expense of threshold estimation, since threshold remains a critical parameter for assessing changes in corticospinal excitability \citep{devanne_input-output_1997,farzan_single-pulse_2014,schicktanz_motor_2014,vucic_clinical_2023,bestmann_use_2024}.
We therefore compared EIG with ML-PEST, EIG focused on threshold (Fig.~\ref{fig_accuracy},~\ref{fig_val}b--e), and variance reduction on threshold (Supplementary Fig.~\ref{sufig_evr}).
EIG performed comparably to these threshold focused designs while matching uniform sampling in predictive accuracy, suggesting it does not trade off threshold estimation for recovering the overall curve.
These improvements hold for both single and multiple-muscle acquisition.
Thus, EIG combines accurate threshold and curve estimation, such that choosing a different design would require more stimuli to reach either criterion.

Our adaptive design can only be as accurate as the underlying model describing neural recruitment, which we previously validated on empirical TMS and SCS data \citep{tyagi_hierarchical_2025}.
Existing adaptive approaches have used Fisher information with a four-parameter logistic function \citep{alavi_optimal_2019}.
These approaches rely on numerical optimization over non-convex search spaces to estimate curve parameters, which is susceptible to suboptimal solutions with small samples and requires repeated random reinitializations to avoid local minima, and provide only point estimates.
In contrast, our approach uses the posterior distribution over curve parameters to quantify estimation uncertainty, and EIG chooses the next stimulus by averaging information gain over this uncertainty rather than relying on a point estimate.
Furthermore, using a recruitment curve function that does not appropriately represent the data can produce systematic errors in parameter estimates and compromise subsequent analyses.
The threshold is neither an explicit parameter nor can it be derived from the logistic functions since they are smooth and cannot capture sharp deflections from the baseline response.
Instead, our rectified-logistic function provides a consistent parametric representation of threshold as a deflection of MEP size from the baseline response.
It outperformed traditional four- and five-parameter logistic functions in predictive performance, as determined through cross-validation on empirical TMS and SCS data (see Fig.~3 of \cite{tyagi_hierarchical_2025}).
Lastly, we do not establish whether EIG is preferable to Fisher information as an acquisition objective; rather, EIG provides a natural information-theoretic criterion within our Bayesian framework.

While we demonstrate feasibility in human TMS experiments, a study comparing EIG with uniform sampling and ML-PEST remains necessary.
However, the present experiments establish the necessary experimental setup and provide a prospective sample size estimate for such a study.
A prospective comparison also introduces experimental variability absent from our simulations.
In particular, the threshold can drift during an experimental session, potentially confounding comparisons between acquisition designs \citep{julkunen_does_2012,tran_expected_2021}.
Such drift may arise from changes in participant state over long sessions, from inadvertent effects of neurostimulation, or from anticipation of the next stimulation event.
To mitigate the latter, we plan to use a minimum interstimulus interval of 5 seconds followed by an exponentially distributed delay, whose memoryless property prevents anticipation of the next stimulus based on elapsed waiting time.
Additional work is needed to characterize the within session changes in threshold and distinguish them from differences attributable to the acquisition design.

Although we do not prescribe a stopping criterion, the posterior distribution available after each update provides a natural basis for adaptive stopping.
For instance, acquisition could terminate once the 95\% HDI of the threshold reaches an acceptable width or its width stabilizes across consecutive stimuli.
The appropriate criterion will depend on the intended application and the precision required for the parameter of interest.
We therefore leave the systematic evaluation of stopping criteria to future work rather than prescribing one criterion for all experiments.

Finally, although we validate our method experimentally for TMS, its deployment in SCS requires additional development.
However, the principal limitations are computational rather than methodological.
SCS spans a larger stimulation range compared to 101 integral values over $[0,100]$\% MSO for TMS, thereby increasing the candidate stimuli over which the acquisition function must be evaluated and consequently the proposal time.
Moreover, SCS may also use interstimulus intervals shorter than the approximately 5 seconds currently required by our implementation.
Faster posterior updates using importance sampling or sequential Monte Carlo could reduce this computational time \citep{moral_sequential_2006,chopin_introduction_2020}.
Nonetheless, the reduction in the total number of stimuli offsets a modest increase in the interstimulus interval, shortening the overall session in SCS.

More broadly, our method is inherently myopic since it optimizes a utility function for one stimulus at a time.
Whether a multistep look-ahead approach would improve acquisition remains an open question.
Policy-based approaches instead optimize decisions over a sequence of experiments.
For instance, deep adaptive design learns an amortized policy by directly optimizing a sequential information objective \citep{foster_deep_2021}.
Such approaches could improve acquisition beyond myopic EIG while shifting much of the computational cost from online acquisition to offline training, making them feasible for SCS.
Further work is needed to determine whether these computational advantages can be achieved without compromising estimation accuracy.
Nonetheless, our generative hierarchical framework provides a basis for training and testing such methods \citep{tyagi_hierarchical_2025}.


In conclusion, the proposed design reduces the experimental burden of acquiring neural recruitment curves.
By allocating samples to informative regions of the curve, it reduces the number of stimuli needed to accurately characterize both the threshold and the overall curve, thereby shortening experimental sessions and reducing inadvertent neuromodulation while enabling neuromuscular parameters to be characterized across multiple muscles simultaneously.
Future work will explore extensions to spatial problems, such as optimizing coil placement in TMS, as well as amortized approaches to facilitate deployment in SCS.

\section{Methods}\label{methods}
\subsection{Bayesian adaptive design}\label{meth_bad}
\subsubsection{Preliminaries}
Let $\mathbb{D} = \left\langle (x_i, y_i) \right\rangle_{i=1}^n$ denote the sequence of $n$ observations collected so far during an experiment, where $x_i \in [0,\infty)$ is the stimulation intensity and $y_i \in (0, \infty)$ is the observed motor-evoked potential (MEP) size at that intensity.
Once we solve the sequential design problem going from $n$ to $n+1$ observations, the same is repeated for each subsequent observation; as such, we omit the dependence on $n$ from the notation.
Let $p(\theta\given\mathbb{D})$ denote the joint posterior over model parameters $\theta$, which is obtained using Markov chain Monte Carlo (MCMC) methods, such as the No-U-Turn sampler (NUTS) \citep{hoffman_no-u-turn_2011}.
In the context of sequential design, the current posterior serves as the prior for choosing the next stimulation intensity $x$.
Hence, to avoid confusion, we refer to $p(\theta\given\mathbb{D})$ as the prior.
Once the next observation $(x, y)$ is available, the updated prior becomes the posterior $p(\theta\given\mathbb{D}, x, y)$.

The model is specified under a likelihood $p(y \given \theta, x)$.
The prior $p(\theta \given \mathbb{D})$ is our knowledge of the model parameters based on the data collected so far.
The evidence $p(y\given\mathbb{D}, x)$ is the predictive distribution of the MEP size $y$ at intensity $x$, obtained by averaging the likelihood over the prior.
Equation~\eqref{eq_terms} summarizes how these relate through Bayes' rule
where we assume the model is completely specified by the model parameters, so that $p\!\left(y \given \mathbb{D}, \theta, x\right) = p\!\left(y\given \theta, x\right)$, and the prior is independent of the design (stimulation intensity) choice, so that $p\!\left(\theta \given \mathbb{D}, x\right) = p\!\left(\theta \given \mathbb{D}\right)$.
\begin{align}
    \underbrace{p(\theta \given \mathbb{D}, x, y)}_{\text{posterior}}
    &=
    \frac{
        \underbrace{p(y \given \theta,x)}_{\text{likelihood}}
        \,
        \underbrace{p(\theta \given \mathbb{D})}_{\text{prior}}
    }{
        \underbrace{p(y \given \mathbb{D}, x)}_{\text{evidence}}
    } \label{eq_terms}
\end{align}

A Monte Carlo (MC) estimator of the evidence is given as,
\begin{align}
    p\!\left(y\given\mathbb{D}, x\right)
    &= \int_\theta p\!\left(y \given \theta, x\right)\, p\!\left(\theta \given \mathbb{D}\right)\, d\theta\notag
    \\[6pt]
    &= \mathbb{E}_{p\!\left(\theta \mid \mathbb{D}\right)}\, p\!\left(y \given \theta,x\right)\notag
    \\[6pt]
    &\approx \frac1{S_1}\sum\limits_{s=1}^{S_1} p\!\left(y\given \theta^{(s)},x\right) \qquad \theta^{(s)} \sim p\!\left(\theta \given \mathbb{D}\right) \label{eq_evidence_mc}
\end{align}

Thus, the sequential design problem is to determine how to optimally choose the next stimulation intensity $x$ based on the data collected so far. The appropriate design criterion depends on the inferential objective---which aspects of the recruitment curve we wish to learn. In the subsequent sections, we present several such objectives.

\subsubsection{Expected information gain (EIG)}\label{meth_eig}
The information gain (IG) from a new observation $(x, y)$ is defined as the Kullback--Leibler (KL) divergence \citep{kullback_information_1951} between the posterior $p\left(\theta \given \mathbb{D}, x, y\right)$ and the prior $p(\theta \given \mathbb{D})$,
\begin{align}
    \mathrm{IG}\!\left(x, y\right) &= \mathrm{KL}\!\left[p\!\left(\theta \given \mathbb{D}, x, y\right) \Vert\, p\!\left(\theta \given \mathbb{D}\right)\right]\notag
    \\[6pt]
    &= \int_\theta \log\!\left(\frac{p\!\left(\theta \given \mathbb{D}, x, y\right)}{p\!\left(\theta \given \mathbb{D}\right)}\right)\, p\!\left(\theta \given \mathbb{D}, x, y\right)\, d\theta \notag
\end{align}

Thus, the information gain measures how much the posterior differs from the prior after receiving a new observation $(x, y)$.
However, when choosing the next intensity $x$, the corresponding MEP size $y$ has not yet been observed.
Instead, the evidence $p(y\given\mathbb{D}, x)$ provides a predictive distribution over the possible MEP sizes $y$ at the candidate intensity $x$.
Therefore, we average the information gain over the evidence to obtain the expected information gain (EIG) \citep{lindley_measure_1956,rainforth_nesting_2018,rainforth_modern_2023},
\begin{align}
    \mathrm{EIG}\!\left(x\right) &= \mathbb{E}_{p\!\left(y\mid \mathbb{D}, x\right)}\, \mathrm{IG}\!\left(x, y\right)\notag
    \\[6pt]
    &= \int_y \int_\theta \log\!\left(\frac{p\!\left(\theta \given \mathbb{D}, x, y\right)}{p\!\left(\theta \given \mathbb{D}\right)}\right)\, p\!\left(\theta \given \mathbb{D}, x, y\right)\, p\!\left(y\given \mathbb{D}, x\right)\, d\theta\, dy\notag
    \\[6pt]
    &= \int_y \int_\theta \log\!\left(\frac{p\!\left(y\given \theta, x\right)}{p\!\left(y\given \mathbb{D}, x\right)}\right)\, p\!\left(y, \theta \given \mathbb{D}, x\right) \, d\theta\, dy \notag
    \\[6pt]
    &= \mathbb{E}_{p\!\left(y, \theta \mid \mathbb{D}, x\right)}\, \log\!\left(\frac{p\!\left(y\given \theta, x\right)}{p\!\left(y\given \mathbb{D}, x\right)}\right)\notag
    \\[6pt]
    &\approx \frac1{S_2}\sum\limits_{s=1}^{S_2} \log\!\left(\frac{p\!\left(y^{(s)}\given \theta^{(s)}, x\right)}{p\!\left(y^{(s)}\given \mathbb{D}, x\right)}\right) \label{eq_eig_mc}
    \\[6pt]
    &\hspace{2.5em} \theta^{(s)}\sim p\!\left(\theta\given \mathbb{D}\right),\quad y^{(s)} \sim p\!\left(y \given \theta^{(s)}, x\right) \notag
\end{align}
Equation~\eqref{eq_eig_mc} gives the nested MC (NMC) estimator of the EIG \citep{huan2013simulation, rainforth_nesting_2018}. It is nested because the denominator $p\!\left(y^{(s)} \given \mathbb{D}, x\right)$ must itself be approximated for each $y^{(s)}$ using \eqref{eq_evidence_mc}, resulting in $\mathcal{O}(S_1 S_2)$ likelihood evaluations for each candidate intensity $x$.


\subsubsection{Focused EIG}
We partition the model parameters as $\theta = \phi \bigsqcup \eta$, i.e., $\theta = \phi \cup \eta$ and $\phi \cap \eta = \varnothing$.
Here, we consider the information gain only about the parameters of interest $\phi$,
\begin{align}
    \mathrm{IG}_{\phi}\!\left(x, y\right) &= \mathrm{KL}\!\left[p\!\left(\phi \given \mathbb{D}, x, y\right) \Vert\, p\!\left(\phi \given \mathbb{D}\right)\right]\notag
    \\[6pt]
    &= \int_\phi \log\!\left(\frac{p\!\left(\phi \given \mathbb{D}, x, y\right)}{p\!\left(\phi \given \mathbb{D}\right)}\right)\, p\!\left(\phi \given \mathbb{D}, x, y\right)\, d\phi\notag    
\end{align}

As before, we average the information gain over the evidence to obtain the focused expected information gain ($\mathrm{EIG}_{\phi}$) \citep{feng_layered_2019},
\begin{align}
    \mathrm{EIG}_{\phi}\!\left(x\right)
    &= \mathbb{E}_{p\!\left(y\mid \mathbb{D}, x\right)}
    \, \mathrm{IG}_{\phi}\!\left(x, y\right)\notag
    \\[6pt]
    &=
    \int_y
    \int_\phi
    \log\!\left(
        \frac{
            p\!\left(\phi \given \mathbb{D},x,y\right)
        }{
            p\!\left(\phi \given \mathbb{D}\right)
        }
    \right)\,
    p\!\left(\phi \given \mathbb{D},x,y\right)\,
    p\!\left(y\given\mathbb{D},x\right)
    \,d\phi\,dy\notag
    \\[6pt]
    &=
    \int_y
    \int_\phi
    \log\!\left(
        \frac{
            p\!\left(y\given \phi, \mathbb{D},x\right)
        }{
            p\!\left(y\given\mathbb{D},x\right)
        }
    \right)\,
    p\!\left(y,\phi \given \mathbb{D},x\right)
    \,d\phi\,dy\notag
    \\[6pt]
    &=
    \int_y
    \int_\phi
    \int_\eta
    \log\!\left(
        \frac{
            p\!\left(y\given \phi, \mathbb{D},x\right)
        }{
            p\!\left(y\given\mathbb{D},x\right)
        }
    \right)\,
    p\!\left(y,\phi,\eta \given \mathbb{D}, x\right)
    \,d\eta\,d\phi\,dy\notag
    \\[6pt]
    &=
    \mathbb{E}_{p\!\left(y,\theta\mid\mathbb{D},x\right)}\,
    \log\!\left(
        \frac{
            p\!\left(y\given\phi,\mathbb{D},x\right)
        }{
            p\!\left(y\given\mathbb{D},x\right)
        }
    \right)\notag
    \\[6pt]
    &\approx
    \frac1{S_2}
    \sum_{s=1}^{S_2}
    \log\!\left(
        \frac{
            p\!\left(y^{(s)}\given\phi^{(s)},\mathbb{D},x\right)
        }{
            p\!\left(y^{(s)}\given\mathbb{D},x\right)
        }
    \right) \label{eq_focused_eig_estimator}
    \\[6pt]
    &\hspace{2.5em}\theta^{(s)} = \left(\phi^{(s)}, \eta^{(s)}\right) \sim p\!\left(\theta\given\mathbb{D}\right),\quad
    y^{(s)} \sim p\!\left(y\given\theta^{(s)},x\right) \notag
\end{align}

The conditional evidence $p\!\left(y^{(s)}\given\phi^{(s)},\mathbb{D},x\right)$ is given as,
\begin{align}
    p\!\left(y^{(s)}\given\phi^{(s)},\mathbb{D},x\right)
    &=
    \int_\eta
    p\!\left(y^{(s)}\given\phi^{(s)},\eta,x\right)\,
    p\!\left(\eta\given\phi^{(s)},\mathbb{D}\right)
    \,d\eta\notag
    \\[6pt]
    &=
    \mathbb{E}_{p\!\left(\eta\mid\phi^{(s)},\mathbb{D}\right)}\,
    p\!\left(y^{(s)}\given\phi^{(s)},\eta,x\right) \notag
    \\[6pt]
    &\approx
    \sum_{r=1}^{S_3}
    \widetilde k_r\!\left(\phi^{(s)}\right)
    \,
    p\!\left(
        y^{(s)}
        \given
        \phi^{(s)}, \eta^{(r)},
        x
    \right) \label{eq_cond_expectation_estimator}
    \\[6pt]
    &\hspace{-7.5em}
    \theta^{(r)} = \left(\phi^{(r)}, \eta^{(r)}\right) \sim p\!\left(\theta \given \mathbb{D}\right),
    \ k_r\!\left(\phi^{(s)}\right)
    = K\!\left(\phi^{(r)}-\phi^{(s)}\right),
    \
    \widetilde k_r\!\left(\phi^{(s)}\right)
    = \frac{k_r\!\left(\phi^{(s)}\right)}{\sum_{\ell=1}^{S_3} k_\ell\!\left(\phi^{(s)}\right)}
    \notag
\end{align}
where $K:\mathbb{R}^{|\phi|}\to[0,\infty)$ is a kernel in $\phi$, and \eqref{eq_cond_expectation_estimator} is the Nadaraya--Watson kernel regression estimator of the conditional expectation \citep{nadaraya_estimating_1964,watson1964smooth}.
Intuitively, it averages over nuisance parameters $\eta^{(r)}$ whose corresponding parameters of interest $\phi^{(r)}$ are close to $\phi^{(s)}$, as determined by the kernel weights $\widetilde{k}_r(\phi^{(s)})$.
These weighted samples approximate the conditional posterior $p\!\left(\eta\given\phi^{(s)},\mathbb{D}\right)$.
Thus, the numerator in \eqref{eq_focused_eig_estimator} requires $\mathcal{O}(S_2S_3)$ likelihood evaluations, resulting in $\mathcal{O}\!\left(S_2(S_1+S_3)\right)$ likelihood evaluations overall.

\subsubsection{Expected variance reduction (EVR)}\label{meth_var-reduction}
A natural objective is to reduce the variance of a function of the model parameters,
$f:\mathbb{R}^{|\theta|}\to\mathbb{R},\ \theta\mapsto f(\theta)$.
For instance, one may define
$f(\theta)\coloneq a$, where $a$ denotes the motor threshold, or
$f(\theta)\coloneq \mathcal{F}(x = 1.2 \cdot a \given \Omega),\, \Omega\subset \theta$, the expected MEP size at
$120\%$ of the motor threshold under the model \eqref{eq_model_start}--\eqref{eq_model_end}.

Consider the reduction in the variance of $f(\theta)$ from the prior to the posterior after receiving a new observation $(x, y)$,
\begin{align}
    \Delta\!\left(x, y\right) &= \mathrm{Var}_{p\!\left(\theta \mid \mathbb{D}\right)} f\!\left(\theta\right) - \mathrm{Var}_{p\!\left(\theta \mid \mathbb{D}, x, y\right)}f\!\left(\theta\right)\notag
\end{align}

As before, we average this under the evidence to obtain the expected variance reduction (EVR) \citep{vanlier_bayesian_2012}.
\begin{align}
    \mathrm{EVR}\!\left(x\right) &= \mathbb{E}_{p\!\left(y\mid \mathbb{D}, x\right)}\ \Delta\!\left(x, y\right)\notag
    \\[6pt]
    &= \mathrm{Var}_{p\!\left(\theta \mid \mathbb{D}\right)} f\!\left(\theta\right) - \mathbb{E}_{p\!\left(y\mid \mathbb{D}, x\right)} \mathrm{Var}_{p\!\left(\theta \mid \mathbb{D}, x, y\right)} f\!\left(\theta\right)\notag
\end{align}

The variance under the prior is given as,
\begin{align}
    \mathrm{Var}_{p\!\left(\theta \mid \mathbb{D}\right)}f\!\left(\theta\right) &= \mathbb{E}_{p\!\left(\theta \mid \mathbb{D}\right)} f\!\left(\theta\right)^2 - \left(\mathbb{E}_{p\!\left(\theta \mid \mathbb{D}\right)} f\!\left(\theta\right)\right)^2\notag
    \\[6pt]
    &\approx \frac1{S_1}\sum\limits_{s=1}^{S_1} f\!\left(\theta^{(s)}\right)^2 - \left(\frac1{S_1}\sum\limits_{s=1}^{S_1} f\!\left(\theta^{(s)}\right)\right)^2 \quad \theta^{(s)}  \sim p\!\left(\theta\given \mathbb{D}\right) \label{eq_var_prior}
\end{align}

The variance under the posterior is given as,
\begin{align}
\mathrm{Var}_{p(\theta \mid \mathbb{D}, x, y)}\,f(\theta)
&= \mathbb{E}_{p(\theta \mid \mathbb{D}, x, y)}\, f(\theta)^2
- \left(\mathbb{E}_{p(\theta \mid \mathbb{D}, x, y)}\,f(\theta)\right)^2\notag
\\[6pt]
&\approx
\sum_{s=1}^{S_1} \widetilde \omega_s\!\left(y\right)\, f\!\left(\theta^{(s)}\right)^2
-
\left(\sum_{s=1}^{S_1} \widetilde \omega_s\!\left(y\right)\, f\!\left(\theta^{(s)}\right)\right)^2\notag
\\[6pt]
&\hspace{-2.5em}\theta^{(s)}  \sim p\!\left(\theta\given \mathbb{D}\right),
\quad \omega_s\!\left(y\right) = p\!\left(y \given \theta^{(s)}, x\right),
\quad
\widetilde \omega_s\!\left(y\right) = \frac{\omega_s\!\left(y\right)}{\sum_{\ell=1}^{S_1} \omega_\ell\!\left(y\right)}
\notag
\end{align}
where $\widetilde\omega_s\!\left(y\right)$ are the self-normalized importance sampling weights obtained when using the prior $p(\theta\given\mathbb{D})$ as the proposal distribution for the posterior $p(\theta\given\mathbb{D},x,y)$ \citep{vanlier_bayesian_2012}.
This follows from \eqref{eq_terms}, since $p\!\left(\theta\given\mathbb{D},x,y\right)\propto p\!\left(y\given\theta,x\right)\,p\!\left(\theta\given\mathbb{D}\right)$, so dividing by the proposal distribution $p\!\left(\theta\given\mathbb{D}\right)$ gives importance weights proportional to the likelihood $p\!\left(y\given\theta,x\right)$.
Finally, the expectation under the evidence is given as,
\begin{align}
\mathbb{E}_{p\!\left(y\mid \mathbb{D}, x\right)} \mathrm{Var}_{p\!\left(\theta \mid \mathbb{D}, x, y\right)}f\!\left(\theta\right)
&= \int_y \mathrm{Var}_{p\!\left(\theta \mid \mathbb{D}, x, y\right)}\!\left[f\!\left(\theta\right)\right] \,p\!\left(y\given \mathbb{D}, x\right)\, dy\notag
\\[6pt]
&= \int_y \mathrm{Var}_{p\!\left(\theta \mid \mathbb{D}, x, y\right)}\!\left[f\!\left(\theta\right)\right]\int\limits_{\theta^\prime\in \mathbb{R}^{|\theta|}} p\!\left(y, \theta^{\prime}\given \mathbb{D}, x\right)\,d\theta^{\prime} \, dy\notag
\\[6pt]
&= \int_y \int\limits_{\theta^\prime\in \mathbb{R}^{|\theta|}} \mathrm{Var}_{p\!\left(\theta \mid \mathbb{D}, x, y\right)}\!\left[f\!\left(\theta\right)\right]\, p\!\left(y, \theta^{\prime}\given \mathbb{D}, x\right)\,d\theta^{\prime} \, dy\notag
\\[6pt]
&= \mathbb{E}_{p\!\left(y,\theta^{\prime}\mid \mathbb{D}, x\right)}\mathrm{Var}_{p\!\left(\theta \mid \mathbb{D}, x, y\right)}\, f\!\left(\theta\right)\notag
\\[6pt]
&\approx
\frac{1}{S_2}\sum\limits_{r=1}^{S_2}
\left[
\sum_{s=1}^{S_1} \widetilde \omega_s\!\left(y^{(r)}\right)\, f\!\left(\theta^{(s)}\right)^2
-
\left(\sum_{s=1}^{S_1} \widetilde \omega_s\!\left(y^{(r)}\right)\, f\!\left(\theta^{(s)}\right)\right)^2
\right]
\notag
\\[6pt]
&\hspace{2.5em}\theta^{\prime(r)}\sim p\!\left(\theta \given \mathbb{D}\right),\ y^{(r)} \sim p\!\left(y\given \theta^{\prime(r)}, x\right) \label{eq_var_post}
\end{align}
thereby requiring $\mathcal{O}(S_1S_2)$ likelihood evaluations.
Supplementary Fig.~\ref{sufig_evr} shows EVR on threshold $(f\!\left(\theta\right)\coloneq a)$ matches EIG focused on threshold $(\phi=\{a\})$ in accuracy while having faster proposal time.

\subsection{Extension to multiple muscles}\label{meth_extension}
More generally, the observation $(x, y)$ consists of MEP sizes $y=(y_m)_{m=1}^M \in (0, \infty)^M$ recorded simultaneously from $M$ muscles at stimulation intensity $x\in[0,\infty)$.
Let $\mathbb{D}_m = \left\langle (x_i, y_{i,m}) \right\rangle_{i=1}^n$ denote the data for muscle $m$, so the collected data partitions as $\mathbb{D} = \bigsqcup_m \mathbb{D}_m$.
Similarly, let $\theta_m$ denote the model parameters for muscle $m$, so the model parameters partition as $\theta = \bigsqcup_m \theta_m$.
Under the following assumptions, we will show the acquisition functions for EIG, focused EIG, and EVR decompose into sums of their corresponding muscle-specific acquisition functions, and hence can be computed in parallel.
We assume the priors are independent across muscles $p(\theta \given \mathbb{D})=\prod_m p(\theta_m \given \mathbb{D}_m)$, and the observations are conditionally independent across muscles given the model parameters, such that $p\!\left(y \given \theta, x\right) = \prod_m p\!\left(y_m \given \theta_m, x\right)$.
It follows that the evidence factorizes as,
\begin{align}
    p\!\left(y\given\mathbb{D}, x\right) &= \int_\theta p\!\left(y \given \theta, x\right)\, p\!\left(\theta \given \mathbb{D}\right)\, d\theta\notag
    \\[6pt]
    &= \int_\theta \prod\limits_m \!\left[p\!\left(y_m \given \theta_m, x\right)\, p\!\left(\theta_m \given \mathbb{D}_m\right)\right]\, d\theta\notag
    \\[6pt]
    &= \prod\limits_m \int_{\theta_m}p\!\left(y_m \given \theta_m, x\right)\, p\!\left(\theta_m \given \mathbb{D}_m\right)\, d\theta_m\notag
    \\[6pt]
    &= \prod\limits_m p\!\left(y_m\given\mathbb{D}_m, x\right) \label{eq_evidence_factor}
\end{align}

Using \eqref{eq_evidence_factor} together with \eqref{eq_terms}, the posterior also factorizes $p\!\left(\theta\given \mathbb{D},x,y\right) = \prod_m p\!\left(\theta_m\given\mathbb{D}_m,x,y_m\right)$.
Now consider,
\begin{align}
    \mathrm{EIG}\!\left(x\right)
    &= \mathbb{E}_{p\!\left(y, \theta \mid \mathbb{D}, x\right)}\, \log\!\left(\frac{p\!\left(y\given \theta, x\right)}{p\!\left(y\given \mathbb{D}, x\right)}\right) \label{eq_eig_decompose_start}
    \\[6pt]
    &= \mathbb{E}_{p\!\left(y, \theta \mid \mathbb{D}, x\right)}\, \log\!\left(\prod_m \frac{p\!\left(y_m\given \theta_m, x\right)}{p\!\left(y_m\given \mathbb{D}_m, x\right)}\right)\notag
    \\[6pt]
    &= \sum_m \mathbb{E}_{p\!\left(y, \theta \mid \mathbb{D}, x\right)}\, \log\!\left(\frac{p\!\left(y_m\given \theta_m, x\right)}{p\!\left(y_m\given \mathbb{D}_m, x\right)}\right)\notag
    \\[6pt]
    &= \sum_m \underbrace{\mathbb{E}_{p\!\left(y_m, \theta_m \mid \mathbb{D}_m, x\right)}\, \log\!\left(\frac{p\!\left(y_m\given \theta_m, x\right)}{p\!\left(y_m\given \mathbb{D}_m, x\right)}\right)}_{\displaystyle \mathrm{EIG}_m\!\left(x\right)}\notag
    \\[6pt]
    &= \sum_m \mathrm{EIG}_m\!\left(x\right) \label{eq_eig_decompose_end}
\end{align}
where $\mathrm{EIG}_m\!\left(x\right)$ is the EIG for muscle $m$ and can be approximated using the MC estimator in \eqref{eq_eig_mc}.

Now assume the muscle-specific parameters partition as $\theta_m = \phi_m \bigsqcup \eta_m$, and define $\phi = \bigsqcup_m \phi_m$ the set of parameters of interest, and $\eta = \bigsqcup_m \eta_m$ the remaining nuisance parameters.
It is easy to see that the conditional posterior factorizes as $p\!\left(\eta\given \phi,\mathbb{D}\right) = \prod_m p\!\left(\eta_m\given\phi_m,\mathbb{D}_m\right)$.
Consequently, the conditional evidence also factorizes as $p\!\left(y\given \phi, \mathbb{D}, x\right) = \prod_m p\!\left(y_m\given \phi_m, \mathbb{D}_m, x\right)$.
An argument similar to \eqref{eq_eig_decompose_start}--\eqref{eq_eig_decompose_end} shows the focused EIG also decomposes across muscles and can be approximated by the sum of muscle-specific focused EIG estimates obtained using \eqref{eq_focused_eig_estimator}.

Finally, let $f\!\left(\theta\right)=\left(f_1\!\left(\theta_1\right), \ldots, f_M\!\left(\theta_M\right)\right)^\top$ denote a function of the model parameters, where $f_m:\mathbb{R}^{|\theta_m|}\to\mathbb{R}$. For instance, one may define $f_m\!\left(\theta_m\right)\coloneq a_m$, where $a_m$ denotes the muscle-specific threshold. We define the reduction in variance as
\begin{align}
    \Delta\!\left(x,y\right) &= \mathrm{tr}\!\left(\mathrm{Cov}_{p\!\left(\theta \mid \mathbb{D}\right)}f\!\left(\theta\right)\right) - \mathrm{tr}\!\left(\mathrm{Cov}_{p\!\left(\theta \mid \mathbb{D}, x, y\right)}f\left(\theta\right)\right)\notag
    \\[6pt]
    &= \sum\limits_m \mathrm{Var}_{p\!\left(\theta \mid \mathbb{D}\right)}f_m\!\left(\theta_m\right) - \sum\limits_m \mathrm{Var}_{p\!\left(\theta \mid \mathbb{D}, x, y\right)}f_m\!\left(\theta_m\right)\notag
    \\[6pt]
    &= \sum\limits_m \underbrace{\left[\mathrm{Var}_{p\!\left(\theta_m \mid \mathbb{D}_m\right)}f_m\!\left(\theta_m\right) - \mathrm{Var}_{p\!\left(\theta_m \mid \mathbb{D}_m, x, y_m\right)}f_m\!\left(\theta_m\right)\right]}_{\Delta_m\!\left(x, y_m\right)}\notag
    \\[6pt]
    &= \sum\limits_m \Delta_m\!\left(x, y_m\right)\notag
\end{align}

Averaging under the evidence gives,
\begin{align}
\mathrm{EVR}\!\left(x\right) &= \mathbb{E}_{p\!\left(y\mid\mathbb{D}, x\right)}\,\Delta\!\left(x, y\right)\notag
\\[6pt]
&= \sum\limits_m \mathbb{E}_{p\!\left(y \mid \mathbb{D}, x\right)} \, \Delta_m\!\left(x, y_m\right)\notag
\\[6pt]
&= \sum\limits_m \underbrace{\mathbb{E}_{p\!\left(y_m \mid \mathbb{D}_m, x\right)} \, \Delta_m\!\left(x, y_m\right)}_{\mathrm{EVR}_m\!\left(x\right)}\notag
\\[6pt]
&= \sum\limits_m \mathrm{EVR}_m\!\left(x\right)\notag
\end{align}
where $\mathrm{EVR}_m\!\left(x\right)$ is the EVR for muscle $m$ and can be approximated using the MC estimators \eqref{eq_var_prior}, \eqref{eq_var_post}.
Supplementary Fig.~\ref{sufig_evr} shows EVR on threshold applied to the acquisition of multiple muscles.

\subsection{Model evaluation}\label{meth_eval}
We model the MEP size $y$ as a rectified-logistic function $\mathcal{F}: \mathbb{R} \to \mathbb{R}^+$ of stimulation intensity $x$ under a gamma (shape--rate) likelihood \eqref{eq_model_start}--\eqref{eq_model_end} \citep{tyagi_hierarchical_2025}.
Supplementary Fig.~\ref{sufig_bayesian_model}a shows the effect of varying the function parameters.
Here, parameter $a$ is the threshold, $b$ controls the growth rate, $g$ is the baseline response of the recording device, $h$ controls the distance between the baseline and saturation ($g+h$), and $v$ controls the location of the inflection point, or point of maximum gradient, whether near the baseline $g$ or saturation ($g+h$).
The parameters $c_1$ and $c_2$ together control the noise around the curve.
Supplementary Fig.~\ref{sufig_bayesian_model}b,c specifies the model used in the update step to estimate the curve parameters from the data acquired in Results~\ref{res_sparse}--\ref{res_shift} (Fig.~\ref{fig_accuracy}--\ref{fig_power}).
For the real TMS experiments in Results~\ref{res_val} (Fig.~\ref{fig_val}a), the model was replaced with a mixture extension to account for outliers (Supplementary Fig.~\ref{sufig_mixture}).
In this case, we turn off the mixture component $(z=0)$ when evaluating the EIG acquisition function.
\begin{align}
    y \given x,\theta &\sim \mathrm{gamma}\!\left(\mu\beta,\beta\right)\label{eq_model_start}
    \\[6pt]
    \mu &= \mathcal{F}\!\left(x\given\Omega\right)\notag
    \\[6pt]
    \beta &= \frac{1}{c_1}+\frac{1}{c_2\mu}\notag
    \\[6pt]
    \mathcal{F}\!\left(x\given\Omega\right) &= g+\max\left\{0,-v+\frac{h+v}{1+\left(\frac{h}{v}\right)e^{-b(x-a)}}\right\}\label{eq_model_end}
    \\[6pt]
    \Omega &= \{a,b,g,h,v\}
    \qquad
    \theta = \Omega\cup\{c_1,c_2\}\subset (0, \infty)\notag
\end{align}

Given any acquired dataset $\mathbb{D}$, we use the model to estimate the parameters with MCMC and obtain current posterior samples $\{\widetilde{\theta}^{(s)}\}_{s=1}^{S}\sim p\!\left(\theta\given\mathbb{D}\right)$.
In simulations, the true model parameters $\theta$ are known.
We evaluate the $L_1$ loss for threshold estimation ($e_a$, absolute error) and curve estimation ($e_y$, integrated absolute error) as
\begin{align}
    e_a
    &= \mathbb{E}_{\widetilde{\theta} \sim p\!\left(\theta\mid\mathbb{D}\right)}
    \left|a-\widetilde{a}\right|
    \notag
    \\[6pt]
    &\approx \frac{1}{S}\sum\limits_{s=1}^S
    \left|a-\widetilde{a}^{(s)}\right|
    \notag
    \\[12pt]
    e_y
    &= \frac{1}{h}
    \mathbb{E}_{\widetilde{\theta} \sim p\!\left(\theta\mid\mathbb{D}\right)}
    \int\limits_{x=0}^{100}
    \left|
        \mathcal{F}\!\left(x\given\Omega_0\right)
        -
        \mathcal{F}\!\left(x\given\widetilde{\Omega}_0\right)
    \right|
    \,dx
    \notag
    \\[6pt]
    &\approx \frac{1}{Sh}\sum\limits_{s=1}^S\int\limits_{x=0}^{100}
    \left|
        \mathcal{F}\!\left(x\given\Omega_0\right)
        -
        \mathcal{F}\!\left(x\given\widetilde{\Omega}_0^{(s)}\right)
    \right|
    \,dx
    \notag
\end{align}
where $\Omega_0=\left(\theta\mysetminus\{c_1,c_2\}\right)\big|_{g=0}$ and $\widetilde{\Omega}_0=\left(\widetilde{\theta}\mysetminus\{\widetilde{c}_1,\widetilde{c}_2\}\right)\big|_{\widetilde{g}=0}$, i.e., we remove the baseline response before comparing the curves.
Since the integral does not have a closed form, we evaluate it numerically using the trapezoidal rule on a grid of 501 points over $[0,100]$\% MSO.
We repeat this evaluation across 2,000 simulations (participants), and Fig.~\ref{fig_accuracy}a,b shows the mean error and standard error of the mean across those repetitions, with $e_a$ labeled as the threshold error and $e_y$ the predictive error.

Both errors are expressed in \% MSO, or the unit of stimulation intensity.
To define an acceptable error, we further define the corresponding relative errors as
\begin{align}
    r_a &= \frac{e_a}{a} \times 100
    \notag
    \\[6pt]
    r_y &= \frac{e_y}{(100 - a)} \times 100
    \notag
\end{align}
i.e., we normalize the predictive error by the activation interval $(100 - a)$ within the range of integration, where we assume the true threshold $a\in (0, 100)$.
For each participant, we follow an acquisition method and calculate its relative threshold error $r_a$ and relative predictive error $r_y$ for number of stimuli from $n=3$ to $n=64$.
To compare acquisition designs by the number of stimuli needed to reach an acceptable error, we estimate where the relative error trajectory crosses that level.
In particular, we model the error trajectory (say $r_q,\, q\in\{a,y\}$) as a piecewise linear function of the number of stimuli in log--log space, which corresponds to a piecewise power decay model in the original space.
For an acceptable relative error $k_q\in(0,\infty)$, the likelihood is
\begin{align}
    \log_2 r_q \given n,\theta_q
    &\sim
    \mathrm{normal}\!\left(\mu_q\!\left(n\right),\omega\right)
    \qquad \theta_q = \{t, \lambda_\mathrm{late}, \rho, \tau, \omega\}
    \notag
    \\[6pt]
    \mu_q\!\left(n\right)
    &=
    \log_2 k_q
    -
    \lambda_{\mathrm{early}}\left(\log_2 n-t\right)
    \notag\\
    &\quad
    -
    \left(\lambda_{\mathrm{late}}-\lambda_{\mathrm{early}}\right)
    \left[
        \max\!\left(0,\log_2 n-\tau\right)
        -
        \max\!\left(0,t-\tau\right)
    \right]
    \notag
    \\[6pt]
    \lambda_{\mathrm{early}}
    &=
    \frac{\lambda_{\mathrm{late}}}{\rho}
    \qquad
    \lambda_{\mathrm{late}}>0
    \qquad
    0<\rho<1
    \notag
\end{align}
The parameters $\lambda_{\mathrm{early}}$ and $\lambda_{\mathrm{late}}$ are the rates of error reduction before and after the change point $n=2^\tau$, respectively.
The constraint $\lambda_{\mathrm{early}}>\lambda_{\mathrm{late}}$ allows the rate of error reduction to decrease as more stimuli are acquired.
By construction, $\mu_q\!\left(2^t\right)=\log_2 k_q$, so the model reaches a median error of $r_q=k_q$ at $n=2^t\in(0,\infty)$ stimuli.
Thus, $t$ determines the number of stimuli (in log space) needed to reach the acceptable error and is the parameter of interest for comparing the acquisition designs.

We used the likelihood above within a hierarchical Bayesian estimation (HBe) model \citep{tyagi_hierarchical_2025} to compare the number of stimuli needed by each design to reach acceptable relative errors of 2\% $(k_a=2)$ and 10\% $(k_y=10)$ in threshold and curve estimation, respectively.
Supplementary Fig.~\ref{sufig_error_delta} specifies the HBe model and Supplementary Fig.~\ref{sufig_error_eg} shows a few resulting estimates.
Specifically, we sampled error trajectories of 16 participants without replacement from the 2,000 simulations and modeled the corresponding values of $t$, with $\mu_0$ describing the population-level location for EIG, and $\mu_\Delta$ the population-level location of paired differences between EIG and each comparison design.
This was repeated 500 times and Fig.~\ref{fig_accuracy}c shows the median posterior estimates of $\mu_0$ for EIG and $(\mu_0+\mu_\Delta)$ for each comparison design across those repetitions.

The simulations and evaluation of true and false positive rates for Fig.~\ref{fig_power} are covered in \cite{tyagi_hierarchical_2025} (see Methods 2.3).
Supplementary Fig.~\ref{sufig_delta} specifies the HBe model used to summarize the location of threshold differences across participants.
The null hypothesis of no difference in threshold between the pre- and post-intervention was rejected if the 95\% highest density interval (HDI) of the location parameter $\mu_\Delta$ excluded zero.

For Fig.~\ref{fig_val}b--e, we repeated the analysis in Fig.~\ref{fig_accuracy} using the eight participants from the real TMS experiments.
Because the ground truth parameters are not available, we instead used the posterior estimated from the 40 samples acquired for each participant by the end of their experimental session.
In particular, we sampled 200 parameter sets from each participant's posterior draws as possible ground truths and evaluated each acquisition design on them.
For Fig.~\ref{fig_val}b,c, we first took the median error over the 200 sets for each participant and then reported the mean and standard error across the eight participants.

For the prospective power analysis in Fig.~\ref{fig_val}d, we used the same HBe model (Supplementary Fig.~\ref{sufig_error_delta}) but varied the participant sample sizes $n\in\{2,4,8,12,16\}$.
For each sample size, we sampled $n$ participants with replacement from the eight participants and paired them with $n$ posterior-draw indices sampled without replacement from $\{1,\ldots,200\}$ to choose the corresponding parameter set from each participant's posterior as ground truth.
The null hypothesis of no difference between EIG and a comparison design in the number of stimuli required to reach an acceptable error was rejected if the 95\% HDI of the location parameter $\mu_\Delta$ excluded zero.
The outcome was a success if the null hypotheses comparing EIG against uniform sampling for threshold estimation and against ML-PEST for curve estimation were rejected simultaneously.
This was repeated 500 times and Fig.~\ref{fig_val}d shows the rejection rate across those repetitions, which reached 80\% power with 12 participants.
Figure~\ref{fig_val}e additionally shows the median posterior estimates of $\mu_0$ for EIG and $(\mu_0+\mu_\Delta)$ for each comparison design across these repetitions with 12 participants.

\subsection{Implementation}
All acquisition methods were initialized with eight samples at zero intensity, followed by three stimuli from uniform sampling.
We do not count the eight samples at zero intensity in the number of stimuli because they do not correspond to actual stimulation events, require no additional effort from the participant or experimenter, and are readily available from the recording device \citep{alavi_optimal_2019}.
EIG does not require initial samples; however, all acquisition methods used the same initial dataset to ensure they begin with the same error.
The update step included four chains, each using 400 warmup samples and 400 post-warmup samples with the NUTS sampler, for a total of $400\times 4 = 1600$ posterior samples.
These samples were also used to evaluate estimation error after each update step (Methods~\ref{meth_eval}).
Throughout acquisition, we used all posterior samples, with $S_1=S_2=S_3=1600$ in \eqref{eq_evidence_mc}--\eqref{eq_var_post}.
The acquisition step was batched across candidate stimuli to avoid GPU memory oversubscription.
The implementation of EIG follows directly from the estimators described above, whereas focused EIG and the comparison designs require a few additional details we note below.
Detailed implementation of each design is available in our Python library \texttt{onMEP}.

\subsubsection{Focused EIG}
Focused EIG used a full covariance Gaussian kernel for the kernel regression estimator in \eqref{eq_cond_expectation_estimator}, with bandwidth determined by the multivariate Silverman rule \citep{nadaraya_estimating_1964,watson1964smooth,silverman_density_2018}.
We describe below how we construct the kernel matrix and obtain the normalized weights in \eqref{eq_cond_expectation_estimator}.
Let $F=|\phi|$ denote the cardinality of the focused set.
We used $S_2 = S_3 \left( = S\right)$ in \eqref{eq_focused_eig_estimator} and \eqref{eq_cond_expectation_estimator}.
Let $U\in\mathbb{R}^{F\times S}$ denote the focused posterior samples mapped to unconstrained space.
We define $\mathbbm{1}\in\mathbb{R}^{S\times1}$ the column vector of ones.
The sample mean and centered samples are
\begin{align*}
    \overline{U}
    = \frac{1}{S}U\mathbbm{1}\in\mathbb{R}^{F\times 1}
    \qquad
    Z
    = U-\overline{U}\mathbbm{1}^T\in\mathbb{R}^{F\times S}
\end{align*}
The covariance matrix $C=\frac{1}{S}ZZ^T \in \mathbb{R}^{F\times F}$ determines the shape and orientation of the kernel.
The Silverman rule gives the bandwidth matrix
\begin{align*}
    h = \left(\frac{S(F+2)}{4}\right)^{-2/(F+4)}
    \qquad
    H = hC \in \mathbb{R}^{F\times F}
\end{align*}
Let $U^{(i)}\in\mathbb{R}^{F\times1}$ denote the $i$th sample (column) of $U$.
The matrix of squared Mahalanobis distances $D = (D_{ij})\in\mathbb{R}^{S\times S}$ is
\begin{align*}
    D_{ij}
    =
    \left(U^{(i)}-U^{(j)}\right)^T
    H^{-1}
    \left(U^{(i)}-U^{(j)}\right)
\end{align*}
The Gaussian kernel matrix $K=(K_{ij}) \in \mathbb{R}^{S\times S}$, and the normalized weights $\widetilde{K}= (\widetilde{K}_{ij})\in\mathbb{R}^{S\times S}$ are then
\begin{align*}
    K_{ij}
    = \exp\!\left(-\frac12 D_{ij}\right) \in (0,1]
    \qquad
    \widetilde{K}_{ij}
    = \frac{K_{ij}}{\sum_{\ell=1}^{S}K_{i\ell}} \in (0,1]
\end{align*}
Thus, $\widetilde{K}_{sr}$ gives the weight $\widetilde{k}_r(\phi^{(s)})$ in \eqref{eq_cond_expectation_estimator}.
The weights are precomputed once per posterior update and reused across candidate stimulation intensities.

\subsubsection{Uniform sampling and ML-PEST}
Uniform sampling was implemented using nested quasi-random sequences in \texttt{SciPy} \citep{2020SciPy-NMeth} so that stimuli remained roughly equispaced for any number of acquired samples from $n=3$ to $n=64$ (Supplementary Fig.~\ref{sufig_quasi}) \citep{halton_efficiency_1960,owen_randomized_2017}.
We used an existing implementation of ML-PEST for single-muscle acquisition (Supplementary Fig.~\ref{sufig_pest}) \citep{wang_three_2023}.
We restricted the baseline response in our simulations to remain below the $50$~$\mu$V cutoff required by ML-PEST.
For multiple-muscle acquisition, we extended ML-PEST using a round-robin strategy across muscles.
This allowed ML-PEST to use all samples recorded simultaneously from multiple muscles at a given intensity, as opposed to applying it to each muscle iteratively, which would discard samples from the remaining muscles and artificially inflate the number of stimuli needed to reduce the average error across muscles.
To our knowledge, such an extension of ML-PEST does not exist, and we make our implementation available in the Python library \texttt{onMEP}.

\subsubsection{Notation}
Where they appear, $\mathrm{normal}\!\left(\mu, \sigma\right)$ is the normal distribution (location--scale), $\mathrm{gamma}\!\left(\alpha, \beta\right)$ the gamma distribution (shape--rate), $\mathrm{truncated\ normal}\!\left(\mu, \sigma\right)$ the truncated-normal on $(0,\infty)$ (location--scale), and $\mathrm{normal}^+\!\left(\sigma\right)$ the half-normal (scale).
$\mathrm{Bernoulli}\!\left(p\right)$ denotes the Bernoulli distribution with success probability $p$, and $\mathrm{Uniform}\!\left(a,b\right)$ the uniform distribution on $[a,b]$.
For convenience, we used the terms statistical power and true positive rate interchangeably when the alternative hypothesis was known to be true.

\subsection{Statistics \& reproducibility}
All models were fit using the No-U-Turn sampler (NUTS) implemented in \texttt{NumPyro} \citep{phan2019composable,bingham2019pyro}.
The Bayesian model used in the update step and the hierarchical Bayesian models used for Bayesian estimation were implemented in \texttt{hbMEP} \citep{tyagi_hierarchical_2025}.
All acquisition functions were implemented in \texttt{JAX}.
The timing benchmarks (Fig.~\ref{fig_timings}) were performed on a machine with an Intel Core i9-13900K CPU for the update step and an NVIDIA RTX 4090 GPU for the proposal step.
Code to reproduce the presented analyses is available on GitHub (see Code availability~\ref{code-availability}).

\subsection{Experimental Validation}
\subsubsection{Participants}
All procedures were reviewed and approved by the Institutional Review Board (IRB) of Columbia University Irving Medical Center (Protocols AAAV6853 and AAAV2760).
The studies were pre-registered at clinicaltrials.gov (NCT07561372, NCT07539025).
Written informed consent was obtained prior to study enrollment, and experimental procedures were conducted in compliance with all applicable institutional and governmental regulations.
Individuals over the age of 18 without neurological injury were eligible for recruitment.
Exclusion criteria included a history of seizures, stroke, traumatic brain injury, neurosurgery, or other neurological diseases; prior adverse reactions to transcranial magnetic stimulation (TMS) or non-invasive neurostimulation; cranial metallic implants or implanted electronic medical devices; current or suspected pregnancy; and the use of medications known to alter cortical excitability or lower seizure threshold.
One participant was left-handed, while the remaining participants were right-handed, as determined by the Edinburgh Handedness questionnaire.

\subsubsection{Transcranial magnetic stimulation}
Transcranial magnetic stimulation (TMS) was delivered using a MagPro X100 stimulator equipped with a C-B70 butterfly coil (MagVenture).
The stimulator was operated in biphasic mode (1.0 ms pulse width), and the coil was positioned over the motor cortex hotspot to elicit an optimal response in the target muscle.
Coil position and orientation were continuously tracked using a stereotactic neuronavigation system (Rogue Research).
Stimulator triggering was controlled via a custom MATLAB interface (The MathWorks) paired with a PCIe-6363 data acquisition board and a BNC-2090A terminal block (National Instruments).
Inter-stimulus intervals (ISIs) were set to 5 seconds.
Participants were seated with their palms facing down on a table and instructed not to move either hand.
They were asked to maintain fixation on a cross positioned in the center of their field of view.

\subsubsection{Electromyographic (EMG) recordings}
To capture both on-target and off-target effects, motor responses were recorded bilaterally from multiple upper extremity muscles.
A common ground electrode was positioned over the lateral epicondyle of the humerus on the targeted arm.
Surface EMG signals were amplified using a Digitimer D360 amplifier and hardware band-pass filtered between 20 and 1900 Hz.
The analog signals were digitized at a sampling rate of 20 kHz using a PCIe-6363 data acquisition board and a BNC-2090A terminal block (National Instruments).
Raw analog signals were stored for subsequent offline analysis, while motor-evoked potentials (MEPs) were concurrently processed in real time in MATLAB R2025b (MathWorks Inc., Natick, MA, USA).
Peak-to-peak MEP size between 15 ms and 75 ms was extracted and transmitted in real time from MATLAB via the Lab Streaming Layer (LSL) API to \texttt{onMEP}.
Prior to data collection, signal quality was validated to ensure that baseline peak-to-peak noise was below 50~$\mu$V.

\subsubsection{Motor hotspot and maximum intensity determination}
TMS was applied to the primary motor cortex (M1) contralateral to the dominant hand.
To determine the motor hotspot, the M1 representation was systematically mapped to identify the coil position that produced the highest MEP sizes in the resting contralateral abductor pollicis brevis (APB) muscle at a constant suprathreshold intensity.
Once identified, this location was saved within the neuronavigation system to guide precise coil placement throughout the experiment.
Subsequently, the participant's tolerable maximum stimulation intensity was determined by incrementally increasing the stimulator output until the individual's comfort threshold was reached.
All participants were able to tolerate the full range of stimulation $[0, 100]$\% MSO.

\subsubsection{Experimental design}
In a pilot with a single participant (Protocol AAAV6853), we sought to demonstrate the efficacy of EIG in a head-to-head comparison against uniform sampling.
A randomized block trial design was used to evaluate the performance of the algorithms implemented as part of \texttt{onMEP}.
The algorithms under test formed a block, and their order was randomized for each trial so that both the experimenter and the participant were blinded to the upcoming condition.
This dataset is used in Fig.~\ref{fig_intro}.

As part of an ongoing study evaluating non-invasive paired brain and peripheral stimulation (Protocol AAAV2760), eight participants additionally had their recruitment curves acquired using \texttt{onMEP} running EIG exclusively.
This dataset is used in Fig.~\ref{fig_val}a.

\section{Additional information}
\subsection{Competing interests}
Jason B. Carmel is a founder and stockholder in BackStop Neural and a scientific advisor and stockholder in SharperSense.
The other authors have nothing to disclose.

\subsection{Funding}
This work was supported by the National Institute of Neurological Disorders and Stroke, National Institutes of Health (awards R01NS124224, 1R03NS141040).

\subsection{Acknowledgements}
We thank Columbia University's Department of Systems Biology IT (NIH: S10OD032433, S10OD021764) for providing computational resources.

\subsection{CRediT authorship contribution statement}
VT: Conceptualization, Methodology, Software, Validation, Formal analysis, Writing - Original Draft, Writing - Review \& Editing.
SSD: Investigation, Data curation, Writing - Review \& Editing.
DT: Methodology, Software, Validation, Writing - Review \& Editing.
JMM: Methodology, Validation, Writing - Review \& Editing.
AG: Methodology, Validation, Writing - Review \& Editing.
JBC: Methodology, Resources, Writing - Review \& Editing, Supervision, Funding acquisition.
JRM: Conceptualization, Methodology, Software, Validation, Formal analysis, Investigation, Resources, Data curation, Writing - Original Draft, Writing - Review \& Editing, Supervision, Project administration, Funding acquisition.

\subsection{Data availability}
The datasets used in this study are available from the corresponding authors upon reasonable request.

\subsection{Code availability}\label{code-availability}
The \texttt{onMEP} Python package for online acquisition of MEP recruitment curves implements the acquisition methods.
The package is available at https://github.com/hbmep/onmep.
Code to reproduce the presented analyses is available at https://github.com/hbmep/onmep-manuscript.

\pagebreak
\bibliography{bibliography}

@misc{hoffman_no-u-turn_2011,
	title = {The {No}-{U}-{Turn} {Sampler}: {Adaptively} {Setting} {Path} {Lengths} in {Hamiltonian} {Monte} {Carlo}},
	shorttitle = {The {No}-{U}-{Turn} {Sampler}},
	doi = {10.48550/arXiv.1111.4246},
	publisher = {arXiv},
	author = {Hoffman, Matthew D. and Gelman, Andrew},
	month = nov,
	year = {2011},
}

@misc{foster_deep_2021,
	title = {Deep {Adaptive} {Design}: {Amortizing} {Sequential} {Bayesian} {Experimental} {Design}},
	shorttitle = {Deep {Adaptive} {Design}},
	doi = {10.48550/arXiv.2103.02438},
	publisher = {arXiv},
	author = {Foster, Adam and Ivanova, Desi R. and Malik, Ilyas and Rainforth, Tom},
	month = mar,
	year = {2021},
}

@article{tyagi_hierarchical_2025,
    title = {Hierarchical {Bayesian} estimation of motor-evoked potential recruitment curves yields accurate and robust estimates},
    author = {Tyagi, Vishweshwar and Murray, Lynda M. and Asan, Ahmet S. and Mandigo, Christopher and Virk, Michael S. and Harel, Noam Y. and Carmel, Jason B. and McIntosh, James R.},
    journal = {Brain Stimulation},
    year = {2025},
    doi = {10.1016/j.brs.2025.09.008},
}

@misc{feng_layered_2019,
	title = {A layered multiple importance sampling scheme for focused optimal Bayesian experimental design},
	url = {http://arxiv.org/abs/1903.11187},
	doi = {10.48550/arXiv.1903.11187},
	number = {{arXiv}:1903.11187},
	publisher = {{arXiv}},
	author = {Feng, Chi and Marzouk, Youssef M.},
	year = {2019},
	langid = {english},
	eprinttype = {arxiv},
	eprint = {1903.11187 [stat]},
}

@misc{rainforth_modern_2023,
	title = {Modern Bayesian Experimental Design},
	url = {http://arxiv.org/abs/2302.14545},
	doi = {10.48550/arXiv.2302.14545},
	number = {{arXiv}:2302.14545},
	publisher = {{arXiv}},
	author = {Rainforth, Tom and Foster, Adam and Ivanova, Desi R. and Smith, Freddie Bickford},
	year = {2023},
	langid = {english},
	eprinttype = {arxiv},
	eprint = {2302.14545 [stat]},
}

@article{vanlier_bayesian_2012,
	title = {A Bayesian approach to targeted experiment design},
	volume = {28},
	issn = {1367-4803},
	url = {https://pmc.ncbi.nlm.nih.gov/articles/PMC3324513/},
	doi = {10.1093/bioinformatics/bts092},
	pages = {1136--1142},
	number = {8},
	journaltitle = {Bioinformatics},
	shortjournal = {Bioinformatics},
	author = {Vanlier, J. and Tiemann, C. A. and Hilbers, P. A. J. and van Riel, N. A. W.},
	year = {2012},
}

@article{lindley_measure_1956,
	title = {On a Measure of the Information Provided by an Experiment},
	volume = {27},
	issn = {0003-4851},
	url = {https://www.jstor.org/stable/2237191},
	pages = {986--1005},
	number = {4},
	journaltitle = {The Annals of Mathematical Statistics},
	publisher = {Institute of Mathematical Statistics},
	author = {Lindley, D. V.},
	year = {1956},
}

@misc{rainforth_nesting_2018,
	title = {On Nesting Monte Carlo Estimators},
	url = {http://arxiv.org/abs/1709.06181},
	doi = {10.48550/arXiv.1709.06181},
	number = {{arXiv}:1709.06181},
	publisher = {{arXiv}},
	author = {Rainforth, Tom and Cornish, Robert and Yang, Hongseok and Warrington, Andrew and Wood, Frank},
	year = {2018},
	langid = {english},
	eprinttype = {arxiv},
	eprint = {1709.06181 [stat]},
}

@article{nadaraya_estimating_1964,
	title = {On Estimating Regression},
	volume = {9},
	url = {https://doi.org/10.1137/1109020},
	doi = {10.1137/1109020},
	pages = {141--142},
	number = {1},
	journaltitle = {Theory of Probability \& Its Applications},
	author = {Nadaraya, E. A.},
	year = {1964},
}

@article{kukke_efficient_2014,
	title = {Efficient and reliable characterization of the corticospinal system using transcranial magnetic stimulation},
	volume = {31},
	issn = {0736-0258},
	url = {https://www.ncbi.nlm.nih.gov/pmc/articles/PMC4744647/},
	doi = {10.1097/WNP.0000000000000057},
	number = {3},
	journal = {Journal of Clinical Neurophysiology},
	author = {Kukke, S. N. and Paine, R. W. and Chao, C. and de Campos, A. C. and Hallett, M.},
	month = jun,
	year = {2014},
	pmid = {24887609},
	pmcid = {PMC4744647},
	pages = {246--252},
}

@article{mcintosh_intraoperative_2023,
	title = {Intraoperative electrical stimulation of the human dorsal spinal cord reveals a map of arm and hand muscle responses},
	volume = {129},
	issn = {0022-3077},
	url = {https://journals.physiology.org/doi/full/10.1152/jn.00235.2022},
	doi = {10.1152/jn.00235.2022},
	number = {1},
	journal = {Journal of Neurophysiology},
	author = {McIntosh, James R. and Joiner, Evan F. and Goldberg, Jacob L. and Murray, Lynda M. and Yasin, Bushra and Mendiratta, Anil and Karceski, Steven C. and Thuet, Earl and Modik, Oleg and Shelkov, Evgeny and Lombardi, Joseph M. and Sardar, Zeeshan M. and Lehman, Ronald A. and Mandigo, Christopher and Riew, K. Daniel and Harel, Noam Y. and Virk, Michael S. and Carmel, Jason B.},
	month = jan,
	year = {2023},
	pages = {66--82},
}

@article{devanne_input-output_1997,
	title = {Input-output properties and gain changes in the human corticospinal pathway},
	volume = {114},
	issn = {1432-1106},
	url = {https://doi.org/10.1007/PL00005641},
	doi = {10.1007/PL00005641},
	language = {en},
	number = {2},
	journal = {Experimental Brain Research},
	author = {Devanne, H. and Lavoie, B. A. and Capaday, C.},
	month = apr,
	year = {1997},
	pages = {329--338},
}

@article{gelman_type_2000,
	title = {Type {S} error rates for classical and {Bayesian} single and multiple comparison procedures},
	volume = {15},
	issn = {1613-9658},
	url = {https://doi.org/10.1007/s001800000040},
	doi = {10.1007/s001800000040},
	language = {en},
	number = {3},
	journal = {Computational Statistics},
	author = {Gelman, Andrew and Tuerlinckx, Francis},
	month = sep,
	year = {2000},
	pages = {373--390},
}

@incollection{farzan_single-pulse_2014,
	address = {New York, NY},
	title = {Single-{Pulse} {Transcranial} {Magnetic} {Stimulation} ({TMS}) {Protocols} and {Outcome} {Measures}},
	isbn = {978-1-4939-0879-0},
	language = {en},
	booktitle = {Transcranial {Magnetic} {Stimulation}},
	publisher = {Springer},
	author = {Farzan, Faranak},
	editor = {Rotenberg, Alexander and Horvath, Jared Cooney and Pascual-Leone, Alvaro},
	year = {2014},
	doi = {10.1007/978-1-4939-0879-0_5},
	pages = {69--115},
}

@article{kruschke_bayesian_2013,
	title = {Bayesian estimation supersedes the t test.},
	volume = {142},
	issn = {1939-2222, 0096-3445},
	url = {http://doi.apa.org/getdoi.cfm?doi=10.1037/a0029146},
	doi = {10.1037/a0029146},
	language = {en},
	number = {2},
	journal = {Journal of Experimental Psychology: General},
	author = {Kruschke, John K.},
	year = {2013},
	pages = {573--603},
}

@book{kruschke_doing_2014,
	address = {Boston},
	edition = {2nd edition},
	title = {Doing {Bayesian} {Data} {Analysis}: {A} {Tutorial} with {R}, {JAGS}, and {Stan}},
	isbn = {978-0-12-405888-0},
	shorttitle = {Doing {Bayesian} {Data} {Analysis}},
	language = {English},
	publisher = {Academic Press},
	author = {Kruschke, John},
	month = nov,
	year = {2014},
    doi = {10.1016/B978-0-12-405888-0.00001-5}
}

@article{vincent_hierarchical_2016,
	title = {Hierarchical {Bayesian} estimation and hypothesis testing for delay discounting tasks},
	volume = {48},
	issn = {1554-3528},
	url = {https://doi.org/10.3758/s13428-015-0672-2},
	doi = {10.3758/s13428-015-0672-2},
	language = {en},
	number = {4},
	journal = {Behavior Research Methods},
	author = {Vincent, Benjamin T.},
	month = dec,
	year = {2016},
	pages = {1608--1620},
}

@article{manson_relationship_2020,
	title = {The relationship between maximum tolerance and motor activation during transcutaneous spinal stimulation is unaffected by the carrier frequency or vibration},
	volume = {8},
	issn = {2051-817X},
	url = {https://www.ncbi.nlm.nih.gov/pmc/articles/PMC7070156/},
	doi = {10.14814/phy2.14397},
	number = {5},
	journal = {Physiological Reports},
	author = {Manson, Gerome A. and Calvert, Jonathan S. and Ling, Jeremiah and Tychhon, Boranai and Ali, Amir and Sayenko, Dimitry G.},
	month = mar,
	year = {2020},
	pmid = {32170844},
	pmcid = {PMC7070156},
	pages = {e14397},
}

@article{antal_no_2004,
	title = {No correlation between oving phosphene and motor thresholds: a transcranial magnetic stimulation study},
	volume = {15},
	issn = {0959-4965},
	shorttitle = {No correlation between oving phosphene and motor thresholds},
	url = {https://journals.lww.com/neuroreport/fulltext/2004/02090/no_correlation_between_oving_phosphene_and_motor.17.aspx},
    doi = {10.1097/00001756-200402090-00017},
	language = {en-US},
	number = {2},
	journal = {NeuroReport},
	author = {Antal, Andrea and Nitsche, Michael A. and Kincses, Tamás Z. and Lampe, Christian and Paulus, Walter},
	month = feb,
	year = {2004},
	pages = {297},
}

@article{hassanzahraee_longer_2019,
	title = {Longer {Transcranial} {Magnetic} {Stimulation} {Intertrial} {Interval} {Increases} {Size}, {Reduces} {Variability}, and {Improves} the {Reliability} of {Motor} {Evoked} {Potentials}},
	volume = {9},
	issn = {2158-0014},
	url = {https://www.liebertpub.com/doi/10.1089/brain.2019.0714},
	doi = {10.1089/brain.2019.0714},
	number = {10},
	journal = {Brain Connectivity},
	author = {Hassanzahraee, Maryam and Zoghi, Maryam and Jaberzadeh, Shapour},
	month = dec,
	year = {2019},
	pages = {770--776},
}

@ARTICLE{2020SciPy-NMeth,
    author  = {Virtanen, Pauli and Gommers, Ralf and Oliphant, Travis E. and
            Haberland, Matt and Reddy, Tyler and Cournapeau, David and
            Burovski, Evgeni and Peterson, Pearu and Weckesser, Warren and
            Bright, Jonathan and {van der Walt}, St{\'e}fan J. and
            Brett, Matthew and Wilson, Joshua and Millman, K. Jarrod and
            Mayorov, Nikolay and Nelson, Andrew R. J. and Jones, Eric and
            Kern, Robert and Larson, Eric and Carey, C J and
            Polat, {\.I}lhan and Feng, Yu and Moore, Eric W. and
            {VanderPlas}, Jake and Laxalde, Denis and Perktold, Josef and
            Cimrman, Robert and Henriksen, Ian and Quintero, E. A. and
            Harris, Charles R. and Archibald, Anne M. and
            Ribeiro, Ant{\^o}nio H. and Pedregosa, Fabian and
            {van Mulbregt}, Paul and {SciPy 1.0 Contributors}},
    title   = {{{SciPy} 1.0: Fundamental Algorithms for Scientific
            Computing in Python}},
    journal = {Nature Methods},
    year    = {2020},
    volume  = {17},
    pages   = {261--272},
    adsurl  = {https://rdcu.be/b08Wh},
    doi     = {10.1038/s41592-019-0686-2},
}

@article{phan2019composable,
    title={Composable Effects for Flexible and Accelerated Probabilistic Programming in NumPyro},
    author={Phan, Du and Pradhan, Neeraj and Jankowiak, Martin},
    journal={arXiv},
    year={2019},
    doi = {10.48550/arXiv.1912.11554}
}

@article{bingham2019pyro,
    author    = {Eli Bingham and
                Jonathan P. Chen and
                Martin Jankowiak and
                Fritz Obermeyer and
                Neeraj Pradhan and
                Theofanis Karaletsos and
                Rohit Singh and
                Paul A. Szerlip and
                Paul Horsfall and
                Noah D. Goodman},
    title     = {Pyro: Deep Universal Probabilistic Programming},
    journal   = {Journal of Machine Learning Research},
    volume    = {20},
    pages     = {28:1--28:6},
    year      = {2019},
    url       = {http://jmlr.org/papers/v20/18-403.html},
    doi       = {10.48550/arXiv.1810.09538}
}

@article{pal_spinal_2022,
	title = {Spinal cord associative plasticity improves forelimb sensorimotor function after cervical injury},
	issn = {0006-8950},
	url = {https://doi.org/10.1093/brain/awac235},
	doi = {10.1093/brain/awac235},
	journal = {Brain},
	author = {Pal, Ajay and Park, HongGeun and Ramamurthy, Aditya and Asan, Ahmet S and Bethea, Thelma and Johnkutty, Meenu and Carmel, Jason B},
	month = sep,
	year = {2022},
	pages = {awac235},
}

@article{greiner_recruitment_2021,
	title = {Recruitment of upper-limb motoneurons with epidural electrical stimulation of the cervical spinal cord},
	volume = {12},
	copyright = {2021 The Author(s)},
	issn = {2041-1723},
	url = {https://www.nature.com/articles/s41467-020-20703-1},
	doi = {10.1038/s41467-020-20703-1},
	language = {en},
	number = {1},
	journal = {Nature Communications},
	author = {Greiner, Nathan and Barra, Beatrice and Schiavone, Giuseppe and Lorach, Henri and James, Nicholas and Conti, Sara and Kaeser, Melanie and Fallegger, Florian and Borgognon, Simon and Lacour, Stéphanie and Bloch, Jocelyne and Courtine, Grégoire and Capogrosso, Marco},
	month = jan,
	year = {2021},
	pages = {435},
}

@article{millet_electrical_2011,
	title = {Electrical stimulation for testing neuromuscular function: from sport to pathology},
	volume = {111},
	issn = {1439-6327},
	shorttitle = {Electrical stimulation for testing neuromuscular function},
	url = {https://doi.org/10.1007/s00421-011-1996-y},
	doi = {10.1007/s00421-011-1996-y},
	language = {en},
	number = {10},
	journal = {European Journal of Applied Physiology},
	author = {Millet, Guillaume Y. and Martin, Vincent and Martin, Alain and Vergès, Samuel},
	month = oct,
	year = {2011},
	pages = {2489--2500},
}

@article{chen_clinical_2008,
	title = {The clinical diagnostic utility of transcranial magnetic stimulation: {Report} of an {IFCN} committee},
	volume = {119},
	issn = {1388-2457},
	shorttitle = {The clinical diagnostic utility of transcranial magnetic stimulation},
	url = {https://www.sciencedirect.com/science/article/pii/S1388245707006189},
	doi = {10.1016/j.clinph.2007.10.014},
	number = {3},
	journal = {Clinical Neurophysiology},
	author = {Chen, Robert and Cros, Didier and Curra, Antonio and Di Lazzaro, Vincenzo and Lefaucheur, Jean-Pascal and Magistris, Michel R. and Mills, Kerry and Rösler, Kai M. and Triggs, William J. and Ugawa, Yoshikazu and Ziemann, Ulf},
	month = mar,
	year = {2008},
	pages = {504--532},
}

@article{koponen_transcranial_2024,
	title = {Transcranial magnetic stimulation input–output curve slope differences suggest variation in recruitment across muscle representations in primary motor cortex},
	volume = {18},
	issn = {1662-5161},
	url = {https://www.frontiersin.org/journals/human-neuroscience/articles/10.3389/fnhum.2024.1310320/full},
	doi = {10.3389/fnhum.2024.1310320},
	language = {English},
	journal = {Frontiers in Human Neuroscience},
	author = {Koponen, Lari M. and Martinez, Miles and Wood, Eleanor and Murphy, David L. K. and Goetz, Stefan M. and Appelbaum, Lawrence G. and Peterchev, Angel V.},
	month = feb,
	year = {2024},
}

@article{stefan_induction_2000,
	title = {Induction of plasticity in the human motor cortex by paired associative stimulation},
	volume = {123},
	issn = {0006-8950},
	url = {https://doi.org/10.1093/brain/123.3.572},
	doi = {10.1093/brain/123.3.572},
	number = {3},
	journal = {Brain},
	author = {Stefan, Katja and Kunesch, Erwin and Cohen, Leonardo G. and Benecke, Reiner and Classen, Joseph},
	month = mar,
	year = {2000},
	pages = {572--584},
}

@article{bunday_motor_2012,
	title = {Motor {Recovery} after {Spinal} {Cord} {Injury} {Enhanced} by {Strengthening} {Corticospinal} {Synaptic} {Transmission}},
	volume = {22},
	issn = {0960-9822},
	url = {https://www.sciencedirect.com/science/article/pii/S0960982212012675},
	doi = {10.1016/j.cub.2012.10.046},
	number = {24},
	journal = {Current Biology},
	author = {Bunday, Karen L. and Perez, Monica A.},
	month = dec,
	year = {2012},
	pages = {2355--2361},
}

@article{macdonald_intraoperative_2013,
	title = {Intraoperative motor evoked potential monitoring – {A} position statement by the {American} {Society} of {Neurophysiological} {Monitoring}},
	volume = {124},
	issn = {1388-2457},
	url = {https://www.sciencedirect.com/science/article/pii/S1388245713010006},
	doi = {10.1016/j.clinph.2013.07.025},
	number = {12},
	journal = {Clinical Neurophysiology},
	author = {MacDonald, D. B. and Skinner, S. and Shils, J. and Yingling, C.},
	month = dec,
	year = {2013},
	pages = {2291--2316},
}

@article{picht_preoperative_2011,
	title = {Preoperative functional mapping for rolandic brain tumor surgery: comparison of navigated transcranial magnetic stimulation to direct cortical stimulation},
	volume = {69},
	issn = {1524-4040},
	shorttitle = {Preoperative functional mapping for rolandic brain tumor surgery},
	doi = {10.1227/NEU.0b013e3182181b89},
	language = {eng},
	number = {3},
	journal = {Neurosurgery},
	author = {Picht, Thomas and Schmidt, Sein and Brandt, Stephan and Frey, Dietmar and Hannula, Henri and Neuvonen, Tuomas and Karhu, Jari and Vajkoczy, Peter and Suess, Olaf},
	month = sep,
	year = {2011},
	pmid = {21430587},
	pages = {581--588; discussion 588},
}

@article{sayenko_spinal_2015,
	title = {Spinal segment-specific transcutaneous stimulation differentially shapes activation pattern among motor pools in humans},
	volume = {118},
	issn = {8750-7587},
	url = {https://journals.physiology.org/doi/full/10.1152/japplphysiol.01128.2014},
	doi = {10.1152/japplphysiol.01128.2014},
	number = {11},
	journal = {Journal of Applied Physiology},
	author = {Sayenko, Dimitry G. and Atkinson, Darryn A. and Dy, Christine J. and Gurley, Katelyn M. and Smith, Valerie L. and Angeli, Claudia and Harkema, Susan J. and Edgerton, V. Reggie and Gerasimenko, Yury P.},
	month = jun,
	year = {2015},
	pages = {1364--1374},
}

@article{hofstoetter_spinal_2021,
	title = {Spinal motor mapping by epidural stimulation of lumbosacral posterior roots in humans},
	volume = {24},
	issn = {2589-0042},
	url = {https://www.sciencedirect.com/science/article/pii/S2589004220311275},
	doi = {10.1016/j.isci.2020.101930},
	language = {en},
	number = {1},
	journal = {iScience},
	author = {Hofstoetter, Ursula S. and Perret, Ivan and Bayart, Aymeric and Lackner, Peter and Binder, Heinrich and Freundl, Brigitta and Minassian, Karen},
	month = jan,
	year = {2021},
	pages = {101930},
}

@article{balbinot_segmental_2023,
	title = {Segmental motor recovery after cervical spinal cord injury relates to density and integrity of corticospinal tract projections},
	volume = {14},
	copyright = {2023 The Author(s)},
	issn = {2041-1723},
	url = {https://www.nature.com/articles/s41467-023-36390-7},
	doi = {10.1038/s41467-023-36390-7},
	language = {en},
	number = {1},
	journal = {Nature Communications},
	author = {Balbinot, Gustavo and Li, Guijin and Kalsi-Ryan, Sukhvinder and Abel, Rainer and Maier, Doris and Kalke, Yorck-Bernhard and Weidner, Norbert and Rupp, Rüdiger and Schubert, Martin and Curt, Armin and Zariffa, Jose},
	month = feb,
	year = {2023},
	pages = {723},
}

@article{arora_utility_2022,
	title = {Utility of transcranial magnetic stimulation in the assessment of spinal cord injury: {Current} status and future directions},
	volume = {3},
	issn = {2673-6861},
	shorttitle = {Utility of transcranial magnetic stimulation in the assessment of spinal cord injury},
	url = {https://www.frontiersin.org/journals/rehabilitation-sciences/articles/10.3389/fresc.2022.1005111/full},
	doi = {10.3389/fresc.2022.1005111},
	language = {English},
	journal = {Frontiers in Rehabilitation Sciences},
	author = {Arora, Tarun and Desai, Naaz and Kirshblum, Steven and Chen, Robert},
	month = oct,
	year = {2022},
}

@article{sharma_repeated_2024,
	title = {Repeated epidural stimulation modulates cervical spinal cord excitability in healthy adult rats},
	volume = {243},
	issn = {1432-1106},
	url = {https://doi.org/10.1007/s00221-024-06965-x},
	doi = {10.1007/s00221-024-06965-x},
	language = {en},
	number = {1},
	journal = {Experimental Brain Research},
	author = {Sharma, Pawan and Rampersaud, Hema and Shah, Prithvi K.},
	month = dec,
	year = {2024},
	pages = {22},
}

@article{wang_three_2023,
	title = {Three novel methods for determining motor threshold with transcranial magnetic stimulation outperform conventional procedures}, issn = {1741-2552},
	url = {http://iopscience.iop.org/article/10.1088/1741-2552/acf1cc},
	doi = {10.1088/1741-2552/acf1cc},
	language = {en},
	journal = {Journal of Neural Engineering},
	author = {Wang, Boshuo and Peterchev, Angel V and Goetz, Stefan M},
	year = {2023},
}

@article{alavi_optimal_2019,
	title = {Optimal {Estimation} of {Neural} {Recruitment} {Curves} {Using} {Fisher} {Information}: {Application} to {Transcranial} {Magnetic} {Stimulation}},
	volume = {27},
	issn = {1558-0210},
	shorttitle = {Optimal {Estimation} of {Neural} {Recruitment} {Curves} {Using} {Fisher} {Information}},
	url = {https://ieeexplore.ieee.org/abstract/document/8705343?casa_token=g_OchCwcZswAAAAA:ebjOCcqV5B-9FxzRM7GsMNLMoY4OtmPjYMOkq6isFDMV22hGKAdLHpMmk9m5b7TTEp8yDXmj9w},
	doi = {10.1109/TNSRE.2019.2914475},
	number = {6},
	journal = {IEEE Transactions on Neural Systems and Rehabilitation Engineering},
	author = {Alavi, Seyed Mohammad Mahdi and Goetz, Stefan M. and Peterchev, Angel V.},
	month = jun,
	year = {2019},
	pages = {1320--1330},
}

@article{kraus_recruitment_2018,
	title = {Recruitment of {Additional} {Corticospinal} {Pathways} in the {Human} {Brain} with {State}-{Dependent} {Paired} {Associative} {Stimulation}},
	volume = {38},
	copyright = {Copyright © 2018 the authors 0270-6474/18/381397-12\$15.00/0},
	issn = {0270-6474, 1529-2401},
	url = {https://www.jneurosci.org/content/38/6/1396},
	doi = {10.1523/JNEUROSCI.2893-17.2017},
	language = {en},
	number = {6},
	journal = {Journal of Neuroscience},
	author = {Kraus, Dominic and Naros, Georgios and Guggenberger, Robert and Leão, Maria Teresa and Ziemann, Ulf and Gharabaghi, Alireza},
	month = feb,
	year = {2018},
	pmid = {29335359},
	pages = {1396--1407},
}

@article{schicktanz_motor_2014,
	title = {Motor threshold predicts working memory performance in healthy humans},
	volume = {1},
	issn = {2328-9503},
	url = {https://www.ncbi.nlm.nih.gov/pmc/articles/PMC4207507/},
	doi = {10.1002/acn3.22},
	number = {1},
	journal = {Annals of Clinical and Translational Neurology},
	author = {Schicktanz, Nathalie and Schwegler, Kyrill and Fastenrath, Matthias and Spalek, Klara and Milnik, Annette and Papassotiropoulos, Andreas and Nyffeler, Thomas and de Quervain, Dominique J-F},
	month = jan,
	year = {2014},
	pmid = {25356384},
	pmcid = {PMC4207507},
	pages = {69--73},
}

@article{bestmann_use_2024,
	title = {On the use of the motor threshold as a dependent variable in {TMS} research},
	volume = {17},
	issn = {1935-861X, 1876-4754},
	url = {https://www.brainstimjrnl.com/article/S1935-861X(24)00115-3/fulltext},
	doi = {10.1016/j.brs.2024.06.009},
	language = {English},
	number = {4},
	journal = {Brain Stimulation: Basic, Translational, and Clinical Research in Neuromodulation},
	author = {Bestmann, Sven},
	month = jul,
	year = {2024},
	pmid = {38944193},
	pages = {780--781},
}

@article{potter-baker_influence_2016,
	title = {Influence of {Corticospinal} {Tracts} from {Higher} {Order} {Motor} {Cortices} on {Recruitment} {Curve} {Properties} in {Stroke}},
	volume = {10},
	issn = {1662-453X},
	url = {https://www.frontiersin.org/journals/neuroscience/articles/10.3389/fnins.2016.00079/full},
	doi = {10.3389/fnins.2016.00079},
	language = {English},
	journal = {Frontiers in Neuroscience},
	author = {Potter-Baker, Kelsey A. and Varnerin, Nicole M. and Cunningham, David A. and Roelle, Sarah M. and Sankarasubramanian, Vishwanath and Bonnett, Corin E. and Machado, Andre G. and Conforto, Adriana B. and Sakaie, Ken and Plow, Ela B.},
	month = mar,
	year = {2016},
}

@article{eginyan_effects_2023,
	title = {Effects of motor stimulation of the tibial nerve on corticospinal excitability of abductor hallucis and pelvic floor muscles},
	volume = {3},
	issn = {2673-6861},
	url = {https://www.ncbi.nlm.nih.gov/pmc/articles/PMC9884817/},
	doi = {10.3389/fresc.2022.1089223},
	journal = {Frontiers in Rehabilitation Sciences},
	author = {Eginyan, Gevorg and Zhou, Xueqing and Williams, Alison M. M. and Lam, Tania},
	month = jan,
	year = {2023},
	pmid = {36726808},
	pmcid = {PMC9884817},
	pages = {1089223},
}

@article{iyer_characterization_2019,
	title = {Characterization of stimulus response curves obtained with transcranial magnetic stimulation from bilateral tibialis anterior muscles post stroke},
	volume = {713},
	issn = {0304-3940},
	url = {https://www.sciencedirect.com/science/article/pii/S0304394019306330},
	doi = {10.1016/j.neulet.2019.134530},
	journal = {Neuroscience Letters},
	author = {Iyer, Pooja C. and Madhavan, Sangeetha},
	month = nov,
	year = {2019},
	pages = {134530},
}

@article{kemlin_redundancy_2019,
	title = {Redundancy {Among} {Parameters} {Describing} the {Input}-{Output} {Relation} of {Motor} {Evoked} {Potentials} in {Healthy} {Subjects} and {Stroke} {Patients}},
	volume = {10},
	issn = {1664-2295},
	url = {https://www.ncbi.nlm.nih.gov/pmc/articles/PMC6537607/},
	doi = {10.3389/fneur.2019.00535},
	journal = {Frontiers in Neurology},
	author = {Kemlin, Claire and Moulton, Eric and Leder, Sara and Houot, Marion and Meunier, Sabine and Rosso, Charlotte and Lamy, Jean-Charles},
	month = may,
	year = {2019},
	pmid = {31178817},
	pmcid = {PMC6537607},
	pages = {535},
}

@article{murphy_lower_2023,
	title = {Lower corticospinal excitability and greater fatigue among people with multiple sclerosis experiencing pain},
	volume = {9},
	issn = {2055-2173},
	url = {https://www.ncbi.nlm.nih.gov/pmc/articles/PMC9830099/},
	doi = {10.1177/20552173221143398},
	number = {1},
	journal = {Multiple Sclerosis Journal - Experimental, Translational and Clinical},
	author = {Murphy, Hannah M. and Fetter, Christopher M. and Snow, Nicholas J. and Chaves, Arthur R. and Downer, Matthew B. and Ploughman, Michelle},
	month = jan,
	year = {2023},
	pmid = {36636581},
	pmcid = {PMC9830099},
	pages = {20552173221143398},
}

@article{filipovic_neurophysiological_2021,
	title = {Neurophysiological {Predictors} of {Response} to {Medication} in {Parkinson}'s {Disease}},
	volume = {12},
	issn = {1664-2295},
	url = {https://www.frontiersin.org/journals/neurology/articles/10.3389/fneur.2021.763911/full},
	doi = {10.3389/fneur.2021.763911},
	language = {English},
	journal = {Frontiers in Neurology},
	author = {Filipović, Saša R. and Kačar, Aleksandra and Milanović, Sladjan and Ljubisavljević, Miloš R.},
	month = nov,
	year = {2021},
}

@article{carson_characterizing_2013,
	title = {Characterizing {Changes} in the {Excitability} of {Corticospinal} {Projections} to {Proximal} {Muscles} of the {Upper} {Limb}},
	volume = {6},
	issn = {1935-861X},
	url = {https://www.sciencedirect.com/science/article/pii/S1935861X13000399},
	doi = {10.1016/j.brs.2013.01.016},
	number = {5},
	journal = {Brain Stimulation},
	author = {Carson, Richard G. and Nelson, Barry D. and Buick, Alison R. and Carroll, Timothy J. and Kennedy, Niamh C. and Cann, Rachel Mac},
	month = sep,
	year = {2013},
	pages = {760--768},
}

@article{kullback_information_1951,
	title = {On Information and Sufficiency},
	volume = {22},
	issn = {0003-4851, 2168-8990},
	url = {https://projecteuclid.org/journals/annals-of-mathematical-statistics/volume-22/issue-1/On-Information-and-Sufficiency/10.1214/aoms/1177729694.full},
	doi = {10.1214/aoms/1177729694},
	pages = {79--86},
	number = {1},
	journaltitle = {The Annals of Mathematical Statistics},
	publisher = {Institute of Mathematical Statistics},
	author = {Kullback, S. and Leibler, R. A.},
	year = {1951},
	langid = {english},
}

@article{harvey_efficient_1997,
	title = {Efficient estimation of sensory thresholds with {ML}-{PEST}},
	url = {https://brill.com/view/journals/sv/11/1/article-p121_15.xml},
	doi = {10.1163/156856897X00159},
	publisher = {Brill},
	author = {Harvey, Lewis O.},
	year = {1997},
	langid = {english},
}

@article{mishory_maximum-likelihood_2004,
	title = {The Maximum-likelihood Strategy for Determining Transcranial Magnetic Stimulation Motor Threshold, Using Parameter Estimation by Sequential Testing Is Faster Than Conventional Methods With Similar Precision},
	volume = {20},
	issn = {1095-0680},
	url = {https://journals.lww.com/ectjournal/fulltext/2004/09000/the_maximum_likelihood_strategy_for_determining.7.aspx},
	pages = {160},
	number = {3},
	journaltitle = {The Journal of {ECT}},
	author = {Mishory, Alexander and Molnar, Christine and Koola, Jejo and Li, Xingbao and Kozel, F. Andrew and Myrick, Hugh and Stroud, Zachary and Nahas, Ziad and George, Mark S.},
	year = {2004},
	langid = {american},
}

@article{pentland_maximum_1980,
	title = {Maximum likelihood estimation: The best {PEST}},
	volume = {28},
	issn = {1532-5962},
	url = {https://doi.org/10.3758/BF03204398},
	doi = {10.3758/BF03204398},
	shorttitle = {Maximum likelihood estimation},
	pages = {377--379},
	number = {4},
	journaltitle = {Perception \& Psychophysics},
	shortjournal = {Perception \& Psychophysics},
	author = {Pentland, Alex},
	year = {1980},
	langid = {english},
}

@article{taylor_pest_1967,
	title = {{PEST}: Efficient Estimates on Probability Functions},
	volume = {41},
	issn = {0001-4966},
	url = {https://doi.org/10.1121/1.1910407},
	doi = {10.1121/1.1910407},
	shorttitle = {{PEST}},
	pages = {782--787},
	number = {4},
	journaltitle = {The Journal of the Acoustical Society of America},
	shortjournal = {J. Acoust. Soc. Am.},
	author = {Taylor, M. M. and Creelman, C. Douglas},
	year = {1967},
}

@incollection{awiszus_tms_2003,
	author = {Awiszus, Friedemann},
	title = {Chapter 2 {TMS} and threshold hunting},
	booktitle = {Supplements to Clinical Neurophysiology},
	volume = {56},
	pages = {13--23},
	publisher = {Elsevier},
	year = {2003},
	url = {https://www.sciencedirect.com/science/chapter/bookseries/abs/pii/S1567424X09702053},
	doi = {10.1016/S1567-424X(09)70205-3},
}

@article{halton_efficiency_1960,
	title = {On the efficiency of certain quasi-random sequences of points in evaluating multi-dimensional integrals},
	volume = {2},
	issn = {0945-3245},
	url = {https://doi.org/10.1007/BF01386213},
	doi = {10.1007/BF01386213},
	pages = {84--90},
	number = {1},
	journaltitle = {Numerische Mathematik},
	shortjournal = {Numer. Math.},
	author = {Halton, J. H.},
	year = {1960},
	langid = {english},
}

@misc{owen_randomized_2017,
	title = {A randomized Halton algorithm in R},
	url = {http://arxiv.org/abs/1706.02808},
	doi = {10.48550/arXiv.1706.02808},
	number = {{arXiv}:1706.02808},
	publisher = {{arXiv}},
	author = {Owen, Art B.},
	year = {2017},
	eprinttype = {arxiv},
	eprint = {1706.02808 [stat.CO]},
}

@article{chaloner1995bayesian,
  author  = {Chaloner, Kathryn and Verdinelli, Isabella},
  title   = {{Bayesian} Experimental Design: A Review},
  journal = {Statistical Science},
  volume  = {10},
  number  = {3},
  pages   = {273--304},
  year    = {1995},
  doi     = {10.1214/ss/1177009939}
}

@article{watson1964smooth,
  author  = {Watson, G. S.},
  title   = {Smooth Regression Analysis},
  journal = {Sankhy{\=a}: The Indian Journal of Statistics, Series A},
  volume  = {26},
  pages   = {359--372},
  year    = {1964}
}

@article{huan2013simulation,
  author  = {Huan, Xun and Marzouk, Youssef M.},
  title   = {Simulation-based optimal {Bayesian} experimental design for nonlinear systems},
  journal = {Journal of Computational Physics},
  volume  = {232},
  number  = {1},
  pages   = {288--317},
  year    = {2013},
  doi     = {10.1016/j.jcp.2012.08.013}
}

@book{silverman_density_2018,
	address = {New York},
	title = {Density Estimation for Statistics and Data Analysis},
	isbn = {978-1-315-14091-9},
	doi = {10.1201/9781315140919},
	pagetotal = {176},
	publisher = {Routledge},
	author = {Silverman, Bernard W.},
	year = {2018},
}

@article{weise_precise_2023,
	title = {Precise motor mapping with transcranial magnetic stimulation},
	volume = {18},
	copyright = {2022 Springer Nature Limited},
	issn = {1750-2799},
	url = {https://www.nature.com/articles/s41596-022-00776-6},
	doi = {10.1038/s41596-022-00776-6},
	language = {en},
	number = {2},
	journal = {Nature Protocols},
	publisher = {Nature Publishing Group},
	author = {Weise, Konstantin and Numssen, Ole and Kalloch, Benjamin and Zier, Anna Leah and Thielscher, Axel and Haueisen, Jens and Hartwigsen, Gesa and Knösche, Thomas R.},
	month = feb,
	year = {2023},
	pages = {293--318},
}

@article{julkunen_does_2012,
	title = {Does second-scale intertrial interval affect motor evoked potentials induced by single-pulse transcranial magnetic stimulation?},
	volume = {5},
	issn = {1935-861X},
	url = {https://www.sciencedirect.com/science/article/pii/S1935861X11001124},
	doi = {10.1016/j.brs.2011.07.006},
	number = {4},
	journal = {Brain Stimulation},
	author = {Julkunen, Petro and Säisänen, Laura and Hukkanen, Taina and Danner, Nils and Könönen, Mervi},
	month = oct,
	year = {2012},
	pages = {526--532},
}

@article{tran_expected_2021,
	title = {Expected {TMS} excites the motor system less effectively than unexpected stimulation},
	volume = {226},
	issn = {1053-8119},
	url = {https://www.sciencedirect.com/science/article/pii/S1053811920310260},
	doi = {10.1016/j.neuroimage.2020.117541},
	journal = {NeuroImage},
	author = {Tran, Dominic M. D. and McNair, Nicolas A. and Harris, Justin A. and Livesey, Evan J.},
	month = feb,
	year = {2021},
	pages = {117541},
}

@article{hernandez-pavon_tms_2023,
	title = {{TMS} combined with {EEG}: {Recommendations} and open issues for data collection and analysis},
	volume = {16},
	issn = {1935-861X},
	shorttitle = {{TMS} combined with {EEG}},
	url = {https://www.sciencedirect.com/science/article/pii/S1935861X23016960},
	doi = {10.1016/j.brs.2023.02.009},
	number = {2},
	journal = {Brain Stimulation},
	author = {Hernandez-Pavon, Julio C. and Veniero, Domenica and Bergmann, Til Ole and Belardinelli, Paolo and Bortoletto, Marta and Casarotto, Silvia and Casula, Elias P. and Farzan, Faranak and Fecchio, Matteo and Julkunen, Petro and Kallioniemi, Elisa and Lioumis, Pantelis and Metsomaa, Johanna and Miniussi, Carlo and Mutanen, Tuomas P. and Rocchi, Lorenzo and Rogasch, Nigel C. and Shafi, Mouhsin M. and Siebner, Hartwig R. and Thut, Gregor and Zrenner, Christoph and Ziemann, Ulf and Ilmoniemi, Risto J.},
	month = mar,
	year = {2023},
	pages = {567--593},
}

@article{vucic_clinical_2023,
	title = {Clinical diagnostic utility of transcranial magnetic stimulation in neurological disorders. {Updated} report of an {IFCN} committee},
	volume = {150},
	issn = {1388-2457},
	url = {https://www.sciencedirect.com/science/article/pii/S1388245723002377},
	doi = {10.1016/j.clinph.2023.03.010},
	journal = {Clinical Neurophysiology},
	author = {Vucic, Steve and Stanley Chen, Kai-Hsiang and Kiernan, Matthew C. and Hallett, Mark and Benninger, David. H. and Di Lazzaro, Vincenzo and Rossini, Paolo M and Benussi, Alberto and Berardelli, Alfredo and Currà, Antonio and Krieg, Sandro M and Lefaucheur, Jean-Pascal and Long Lo, Yew and Macdonell, Richard A and Massimini, Marcello and Rosanova, Mario and Picht, Thomas and Stinear, Cathy M and Paulus, Walter and Ugawa, Yoshikazu and Ziemann, Ulf and Chen, Robert},
	month = jun,
	year = {2023},
	pages = {131--175},
}

@article{yuasa_systematic_2022,
	title = {Systematic determination of muscle groups and optimal stimulation intensity for simultaneous {TMS} mapping of multiple muscles in the upper limb},
	volume = {10},
	issn = {2051-817X},
	url = {https://onlinelibrary.wiley.com/doi/abs/10.14814/phy2.15527},
	doi = {10.14814/phy2.15527},
	language = {en},
	number = {23},
	journal = {Physiological Reports},
	author = {Yuasa, Akiko and Uehara, Shintaro and Sawada, Yusuke and Otaka, Yohei},
	year = {2022},
	pages = {e15527},
}

@article{nazarova_mapping_2021,
	title = {Mapping of multiple muscles with transcranial magnetic stimulation: absolute and relative test–retest reliability},
	volume = {42},
	issn = {1097-0193},
	shorttitle = {Mapping of multiple muscles with transcranial magnetic stimulation},
	url = {https://onlinelibrary.wiley.com/doi/abs/10.1002/hbm.25383},
	doi = {10.1002/hbm.25383},
	language = {en},
	number = {8},
	journal = {Human Brain Mapping},
	author = {Nazarova, Maria and Novikov, Pavel and Ivanina, Ekaterina and Kozlova, Ksenia and Dobrynina, Larisa and Nikulin, Vadim V.},
	year = {2021},
	pages = {2508--2528},
}

@article{de_freitas_spinal_2026,
	title = {Spinal cord stimulation for upper limb motor function in people with chronic post-stroke hemiparesis: a feasibility trial},
	volume = {32},
	copyright = {2026 The Author(s)},
	issn = {1546-170X},
	shorttitle = {Spinal cord stimulation for upper limb motor function in people with chronic post-stroke hemiparesis},
	url = {https://www.nature.com/articles/s41591-026-04435-1},
	doi = {10.1038/s41591-026-04435-1},
	language = {en},
	number = {7},
	journal = {Nature Medicine},
	publisher = {Nature Publishing Group},
	author = {de Freitas, Roberto M. and Bhatia, Shovan and Sorensen, Erynn and Verma, Nikhil and Carranza, Erick and Ensel, Scott and Borda, Luigi and Boos, Amy and Goldsmith, Jeff and Fisher, Lee E. and Fields, Daryl P. and Powell, Marc P. and Gordon, Shane and Balzer, Jeffrey and Friedlander, Robert M. and Wittenberg, George F. and Gerszten, Peter C. and Krakauer, John W. and Pirondini, Elvira and Weber, Douglas J. and Capogrosso, Marco},
	month = jul,
	year = {2026},
	pages = {2579--2590},
}

@article{suzuki_reciprocal_2012,
	title = {Reciprocal changes in input–output curves of motor evoked potentials while learning motor skills},
	volume = {1473},
	issn = {0006-8993},
	url = {https://www.sciencedirect.com/science/article/pii/S0006899312012322},
	doi = {10.1016/j.brainres.2012.07.043},
	journal = {Brain Research},
	author = {Suzuki, Makoto and Kirimoto, Hikari and Onishi, Hideaki and Yamada, Sumio and Tamaki, Hiroyuki and Maruyama, Atsuo and Yamamoto, Jun-ichi},
	month = sep,
	year = {2012},
	pages = {114--123},
}

@article{murray_timing-dependent_2025,
	title = {Timing-dependent synergies between noninvasive motor cortex and spinal cord stimulation in chronic cervical spinal cord injury},
	volume = {180},
	issn = {1388-2457},
	url = {https://www.sciencedirect.com/science/article/pii/S1388245725012246},
	doi = {10.1016/j.clinph.2025.2111372},
	journal = {Clinical Neurophysiology},
	author = {Murray, Lynda M. and McIntosh, James R. and Goldsmith, Jacob A. and Wu, Yu-Kuang and Liu, Mingxiao and Sanford, Sean P. and Joiner, Evan F. and Mandigo, Christopher and Tyagi, Vishweshwar and Virk, Michael S. and Carmel, Jason B. and Harel, Noam Y.},
	month = dec,
	year = {2025},
	pages = {2111372},
}

@misc{pascual-leone_electrode_2026,
	title = {Electrode position, distance, size, and orientation determine efficacy of cervical epidural stimulation to recruit forelimb muscles in rats},
	copyright = {© 2026, Posted by openRxiv. This pre-print is available under a Creative Commons License (Attribution-NonCommercial-NoDerivs 4.0 International), CC BY-NC-ND 4.0, as described at http://creativecommons.org/licenses/by-nc-nd/4.0/},
	url = {https://www.biorxiv.org/content/10.1101/2025.09.05.674051v2},
	doi = {10.1101/2025.09.05.674051},
	language = {en},
	urldate = {2026-09-29},
	publisher = {bioRxiv},
	author = {Pascual-Leone, Andrés and Tyagi, Vishweshwar and Asan, Ahmet S. and Rocha-Flores, Pedro E. and Rodriguez-Lopez, Ovidio and Voit, Walter and McIntosh, James R. and Carmel, Jason B.},
	month = apr,
	year = {2026},
}

@article{moral_sequential_2006,
	title = {Sequential {Monte} {Carlo} samplers},
	volume = {68},
	issn = {1467-9868},
	url = {https://rss.onlinelibrary.wiley.com/doi/10.1111/j.1467-9868.2006.00553.x},
	doi = {10.1111/j.1467-9868.2006.00553.x},
	language = {en},
	number = {3},
	journal = {Journal of the Royal Statistical Society: Series B (Statistical Methodology)},
	publisher = {John Wiley \& Sons, Ltd},
	author = {Moral, Pierre Del and Doucet, Arnaud and Jasra, Ajay},
	month = jun,
	year = {2006},
	pages = {411--436},
}

@book{chopin_introduction_2020,
	address = {Cham},
	series = {Springer {Series} in {Statistics}},
	title = {An {Introduction} to {Sequential} {Monte} {Carlo}},
	copyright = {https://www.springer.com/tdm},
	isbn = {978-3-030-47844-5 978-3-030-47845-2},
	url = {https://link.springer.com/10.1007/978-3-030-47845-2},
	doi = {10.1007/978-3-030-47845-2},
	language = {en},
	publisher = {Springer International Publishing},
	author = {Chopin, Nicolas and Papaspiliopoulos, Omiros},
	year = {2020},
}

\pagebreak

\setcounter{figure}{0}
\renewcommand{\thefigure}{S\arabic{figure}}
\renewcommand{\theHfigure}{S\arabic{figure}}
\renewcommand{\figurename}{Supplementary Fig.}

\section*{Supplementary information}

\begin{figure}[!htbp]
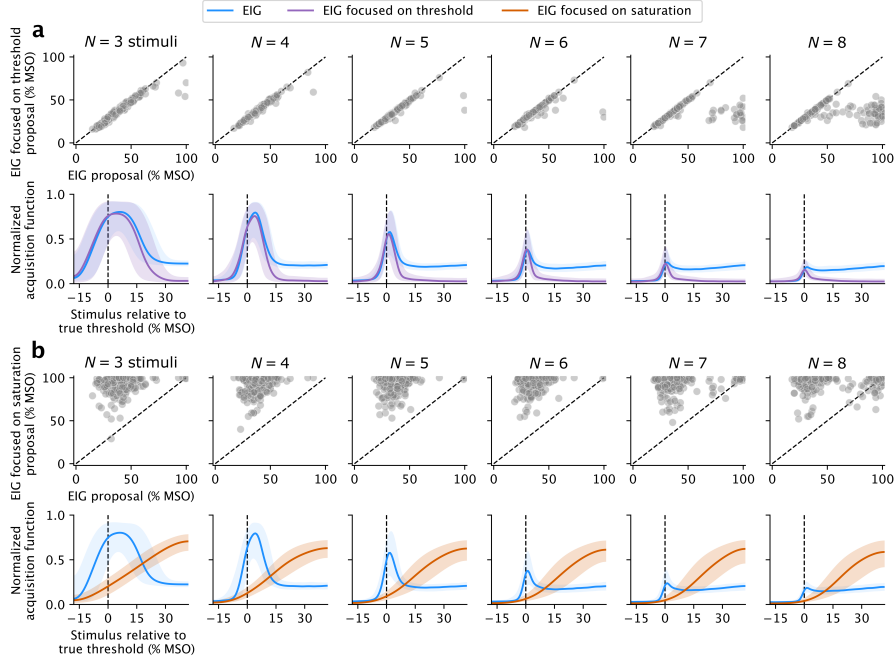

    \centering
    \maybeincludegraphics[width=\figwb\textwidth]{pdf/traj.\figext}
    \caption{
    \textbf{EIG proposals match EIG focused on threshold during early acquisition.}
    \textbf{(a)} Comparison of stimuli proposed by EIG with those that would have been proposed by EIG focused on threshold at the same acquisition step across 200 simulations of EIG for $N\in[3, 8]$ stimuli.
    The dashed diagonal line indicates identical proposals.
    Bottom: corresponding acquisition values as a function of stimulus intensity centered at true threshold (dashed vertical line).
    Acquisition functions are normalized to $[0,1]$ within each simulation run, showing that the relative information gain decreases as more samples are acquired.
    Lines indicate the median and shaded regions the interquartile range across simulations.
    \textbf{(b)} Same as (a), comparing EIG with EIG focused on saturation, whose proposals and acquisition functions do not agree.
    }\label{sufig_traj}
\end{figure}

\begin{figure}[!htbp]
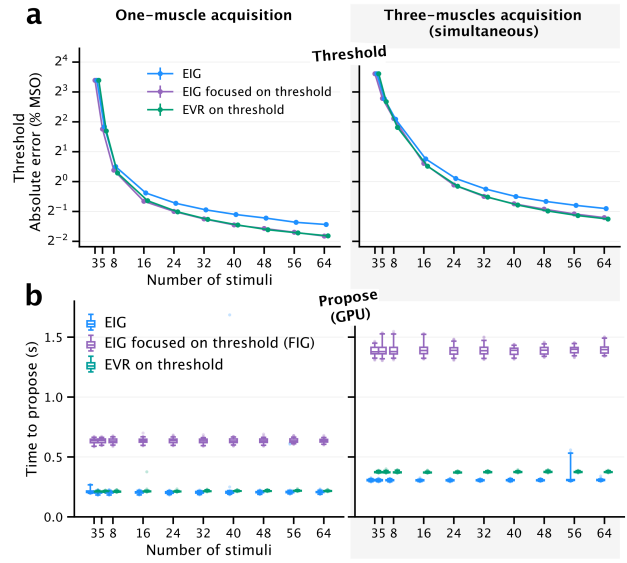

    \centering
    \maybeincludegraphics[width=\figwa\textwidth]{pdf/evr.\figext}
    \caption{
    \textbf{EVR matches EIG focused on threshold in accuracy while having faster proposal time.}
    \textbf{(a)} Mean absolute error in threshold estimation when acquiring a single target muscle (left) and three muscles simultaneously (right).
    \textbf{(b)} Computation time for the proposal step.
    }\label{sufig_evr}
\end{figure}

\begin{figure}[!htbp]
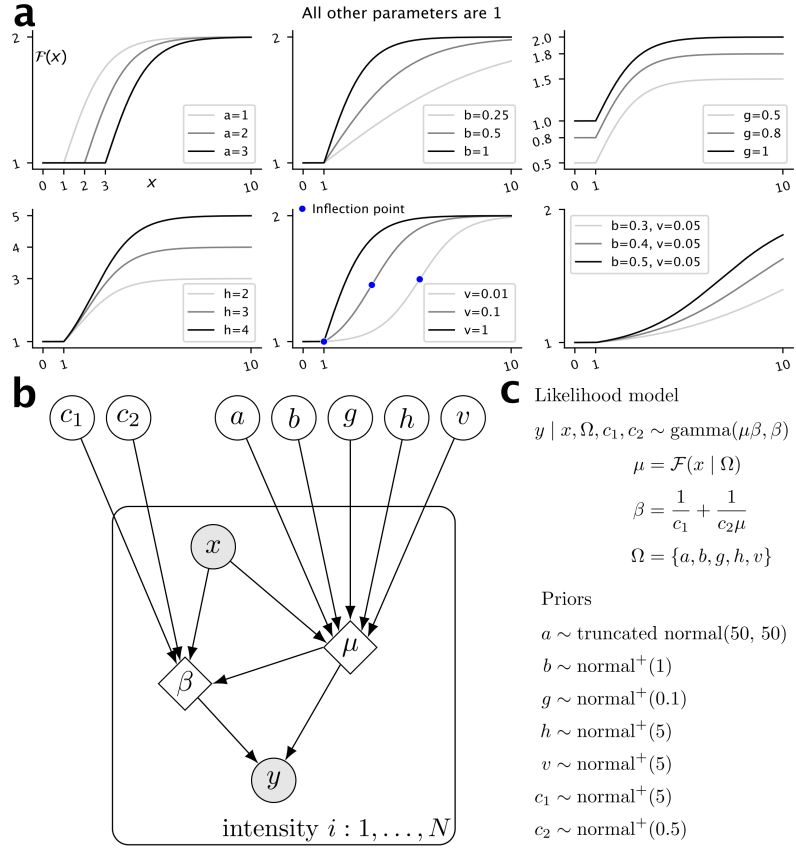

    \centering
    \maybeincludegraphics[width=.80\textwidth]{pdf/bayesian_model.\figext}
    \caption{
    \textbf{Bayesian model specification for TMS data.}
    \textbf{(a)} Effect of varying parameters of the rectified-logistic function.
    \textbf{(b)} Graphical model.
    Circular nodes represent random variables.
    Filled circular nodes represent observed data.
    Diamonds represent deterministic variables.
    Arrows indicate that a child node is informed by the distribution of its parent node.
    Plates denote re-instantiation of nodes.
    \textbf{(c)} Model specification with priors for TMS data.
    Here, $\mathcal{F}$ is the rectified-logistic function.
    }\label{sufig_bayesian_model}
\end{figure}

\begin{figure}[!htbp]
    \centering
    \maybeincludegraphics[width=.99\textwidth]{pdf/mixture.\figext}
    \caption{
    \textbf{Mixture extension of the Bayesian model.}
    }\label{sufig_mixture}
\end{figure}

\begin{figure}[!htbp]
    \centering
    \maybeincludegraphics[width=\figwa\textwidth]{pdf/error_delta.\figext}
    \caption{
    \textbf{Paired comparison model for the number of stimuli needed to reach acceptable error.}
    Here $d=1$ represents EIG, and $d > 1$ may represent EIG focused on threshold, ML-PEST, or uniform sampling.
    A priori, the model assumes no difference in the number of stimuli needed to reach an acceptable median error $k_q$, as indicated by the weak prior on $\mu_\Delta$ centered at zero.
    }\label{sufig_error_delta}
\end{figure}

\begin{figure}[!htbp]
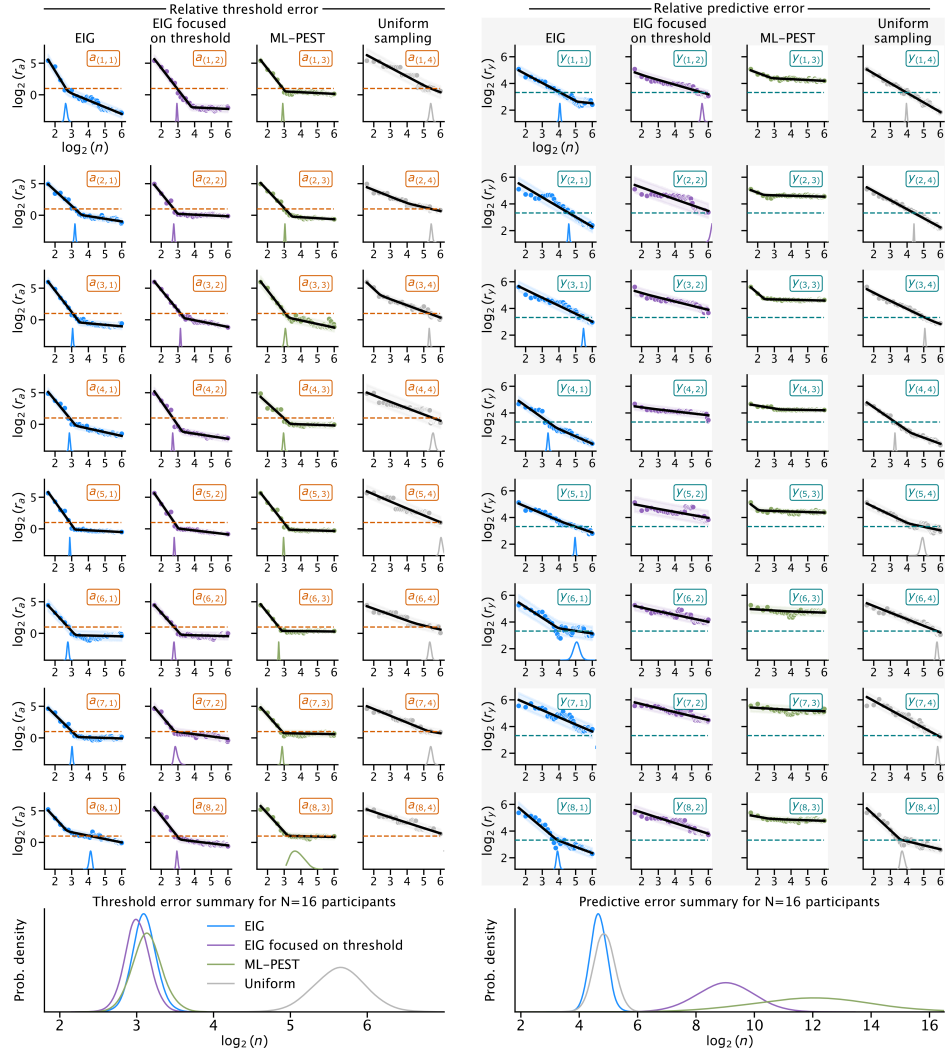

    \centering
    \maybeincludegraphics[width=.95\textwidth]{pdf/error.\figext}
    \caption{
    \textbf{Example piecewise linear functions estimated using the paired comparison model comparing the number of stimuli needed to reach acceptable error across designs.}
    Each row, except the last, shows one participant, with columns corresponding to EIG, EIG focused on threshold, ML-PEST, and uniform sampling.
    Only the first eight of the 16 participants are shown.
    Left: relative threshold error $r_a$, with the orange horizontal dashed line indicating $\log_2 k_a$, corresponding to acceptable error of $2$\% $(k_a=2)$.
    Right: relative predictive error $r_y$, with the teal horizontal dashed line indicating $\log_2 k_y$, corresponding to acceptable error of $10$\% $(k_y=10)$.
    The number of stimuli $n$ and relative errors are shown in log--log space, with dots indicating the relative errors and black lines the estimated piecewise linear functions.
    Inset: posterior distribution of the number of stimuli needed to reach the corresponding acceptable error.
    A missing or truncated posterior indicates that the estimated number of stimuli needed to reach acceptable error extends beyond the range shown.
    Bottom row: posterior distributions of population-level location $\mu_0$ for EIG and $(\mu_0 + \mu_{\Delta})$ for each comparison design, summarizing the number of stimuli needed to reach acceptable error across 16 participants.
    }\label{sufig_error_eg}
\end{figure}

\begin{figure}[!htbp]
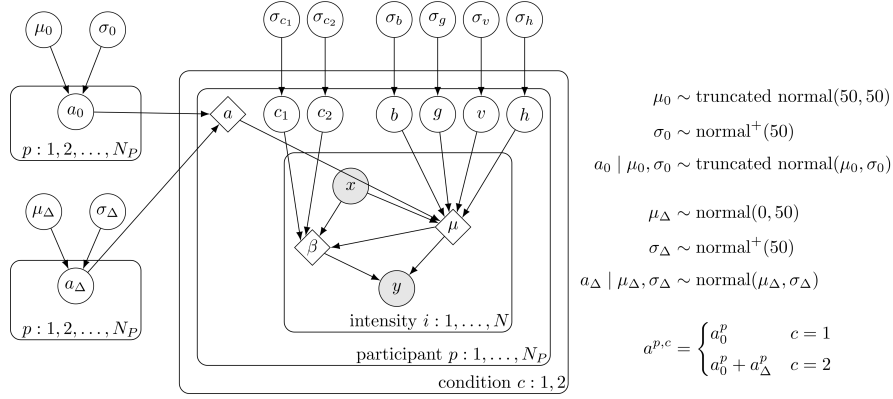

    \centering
    \maybeincludegraphics[width=\figwb\textwidth]{pdf/delta.\figext}
    \caption{
    \textbf{Paired comparison model for detecting a shift in threshold between pre- and post-intervention phases.}
    Here $c=1$ and $c=2$ represent pre- and post-intervention conditions, respectively.
    A priori, the model assumes no shift, as indicated by the weak prior on $\mu_\Delta$ centered at zero.
    }\label{sufig_delta}
\end{figure}

\begin{figure}[!htbp]
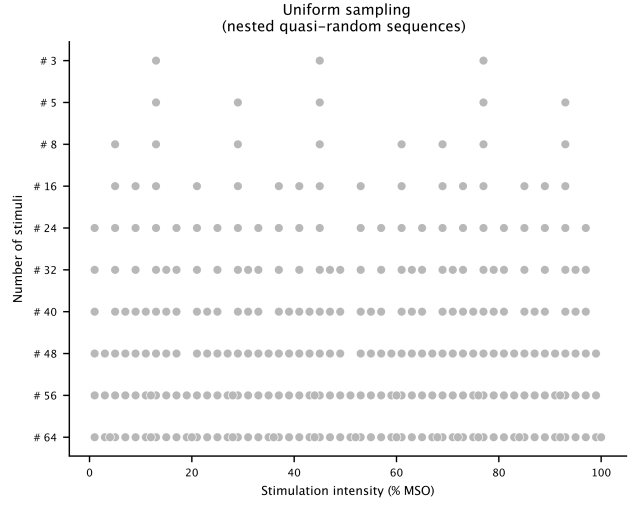

    \centering
    \maybeincludegraphics[width=\figwa\textwidth]{pdf/quasi.\figext}
    \caption{
    \textbf{Uniform sampling using nested quasi-random sequence.}
    Stimuli remain approximately evenly distributed over $[0,100]$\% MSO at any number of acquired samples from $n=3$ to $n=64$, with each larger set containing all samples from the smaller sets.
    }\label{sufig_quasi}
\end{figure}

\begin{figure}[!htbp]
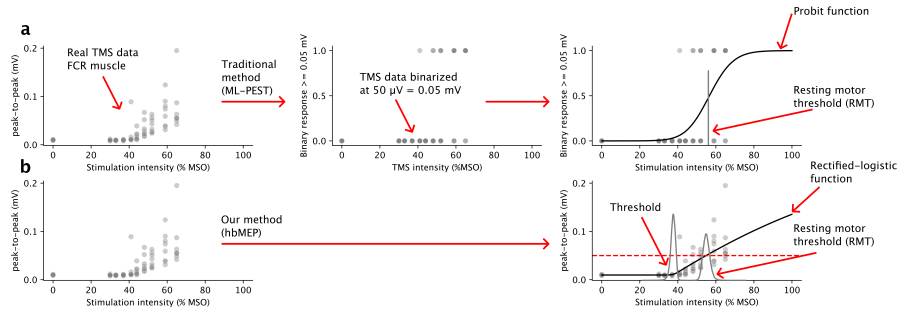

    \centering
    \maybeincludegraphics[width=\figwb\textwidth]{pdf/pest.\figext}
    \caption{
    \textbf{ML-PEST is designed for the resting motor threshold.}
    \textbf{(a)} ML-PEST binarizes the MEP sizes using the conventional $50$~$\mu$V cutoff and fits a probit function to the resulting binary responses.
    The resting motor threshold (RMT) is the stimulation intensity at which the fitted probability of observing an MEP larger than $50$~$\mu$V is $0.5$.
    ML-PEST provides a point estimate of the RMT.
    \textbf{(b)} In contrast, hbMEP directly models the observed MEP sizes using the rectified-logistic function and defines the threshold as the stimulation intensity at which the expected MEP size deflects from the baseline response.
    The RMT can additionally be obtained from hbMEP as a postprocessing step, which agrees with the ML-PEST estimate.
    However, the threshold cannot be estimated using ML-PEST.
    Additionally, ML-PEST fails when the baseline response exceeds the fixed $50$~$\mu$V cutoff, whereas hbMEP can accommodate different baseline response levels since it estimates the baseline within the same model.
    }\label{sufig_pest}
\end{figure}

\end{document}